\documentclass[10pt,aps,prl,reprint,showpacs,superscriptaddress,floats,floatfix,amsmath,amssymb]{revtex4-1}
\usepackage[pdfpagemode=None]{hyperref}
\usepackage{graphicx}
\usepackage{xcolor}

\begin{document}
\title{One-Dimensional Transport of Bosons  between Weakly Linked Reservoirs}
\date{7 January   2014; revised 13 February 2014}
\author{D. P. Simpson} \affiliation{School of Physics and Astronomy, University
  of Birmingham, Edgbaston, Birmingham, B15 2TT, United Kingdom}
\author{D. M. Gangardt} \affiliation{School of Physics and Astronomy,
  University of Birmingham, Edgbaston, Birmingham, B15 2TT, United Kingdom}
\author{I. V. Lerner} \affiliation{School of Physics and Astronomy, University
  of Birmingham, Edgbaston, Birmingham, B15 2TT, United Kingdom}
  \author{P. Kr\"{u}ger}\affiliation{Midlands Ultracold Atom Research Centre, School of Physics and Astronomy,
University of Nottingham, University Park, Nottingham, NG7 2RD, UK}
\begin{abstract}
  We study a flow of ultracold bosonic atoms through a one-dimensional
  channel that connects two macroscopic three-dimensional reservoirs of Bose-condensed atoms via  weak links implemented as potential barriers between each of the reservoirs and
  the channel. We consider reservoirs at equal chemical potentials so that a superflow of the quasi-condensate through the channel is driven purely by a phase difference, $2\Phi$,  imprinted  between the reservoirs.  We find that the superflow never has the standard Josephson form $\sim \sin 2\Phi $. Instead, the superflow discontinuously flips direction at $2\Phi =\pm\pi$ and  has metastable branches. We show that these features are robust and not smeared by fluctuations or phase slips. We describe a possible experimental setup for observing these phenomena.
\end{abstract}
\maketitle

 \newcommand{\dphi}{\widetilde \phi}
\newcommand{\dd}{\operatorname{d}}
\newcommand{\sgn}{\operatorname{sgn}}

Recent advances in trapping and manipulating ultracold gases have enabled experimental observations of a variety of new transport phenomena in quasi- one-dimensional (1D) cold atom systems  \cite{Billy:2008,*Roati:2008,Levy2007,Kohl2009,*Minardi:12,Esslinger2,Esslinger1,TunWeakLink,Tanzi:2013}, complementary to those extensively studied in condensed matter physics. Correlation effects play a crucial role in the behavior of 1D systems and a lot of theoretical effort has been concentrated on the  understanding  of such effects in ultracold gases (see for reviews \cite{CazalillaRMP:11,ImSchGl}).

\begin{figure}[b]
  \includegraphics[width=0.95\columnwidth]{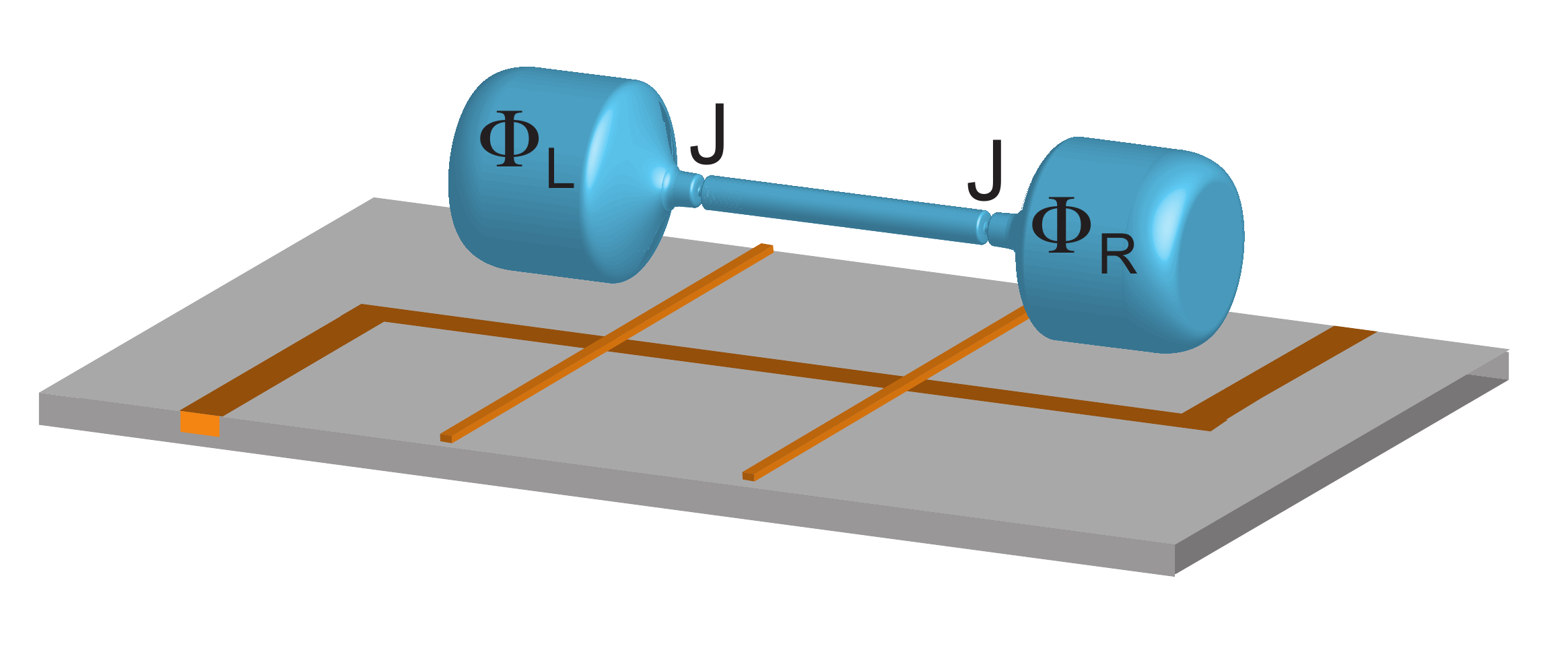}
  \caption{(Color  online) A sketch of two 3D BEC reservoirs connected by a 1D channel via weak tunneling links. A simplified illustration of an atom chip creating an appropriate magnetic trapping configuration is also shown.}
  \label{fig:1}
\end{figure}
In particular,   correlation effects are responsible for a drastic modification of tunneling into a 1D channel and of a 1D flow across a single imperfection, impurity or weak link, as has been shown in numerous theoretical \cite{KaneFis:92a,*Kane-Fisher,MatYueGlaz:93,*FurusakiNagaosa:93b,%
*FabrizioGogolin:95,FurMatv:02,*NazGlaz:03,*PolGorn:03,LYY:08,*GB:10} and experimental \cite{Bockrath:99,*Bockrath:01,Yao:99,Auslaender:02,*Kim:06,Levy:06,*Levy:2012} studies of electronic transport in systems such as semiconductor quantum wires or carbon nanotubes.  A geometry where these types of phenomena can be observed for ultracold atomic systems has rapidly attracted theoretical interest \cite{Paul2007,Gutman-Gefen-Mirlin2012,Reimann2013} and has been recently realized experimentally \cite{Esslinger2,Esslinger1}  by connecting 3D fermionic reservoirs via a 1D channel. A similar experiment with ultracold \emph{bosons}   would lead to the intriguing opportunity to explore   coherent 1D   transport focusing on   features without a direct analogy in condensed matter systems.

In this Letter we study a 1D flow of degenerate ultracold bosons driven by a phase difference between two macroscopic Bose--Einstein condensates (BEC), which are weakly connected by a 1D channel via two tunneling barriers (see Fig~\ref{fig:1}).

\begin{figure}[b]
\centering
\includegraphics[width=\columnwidth]{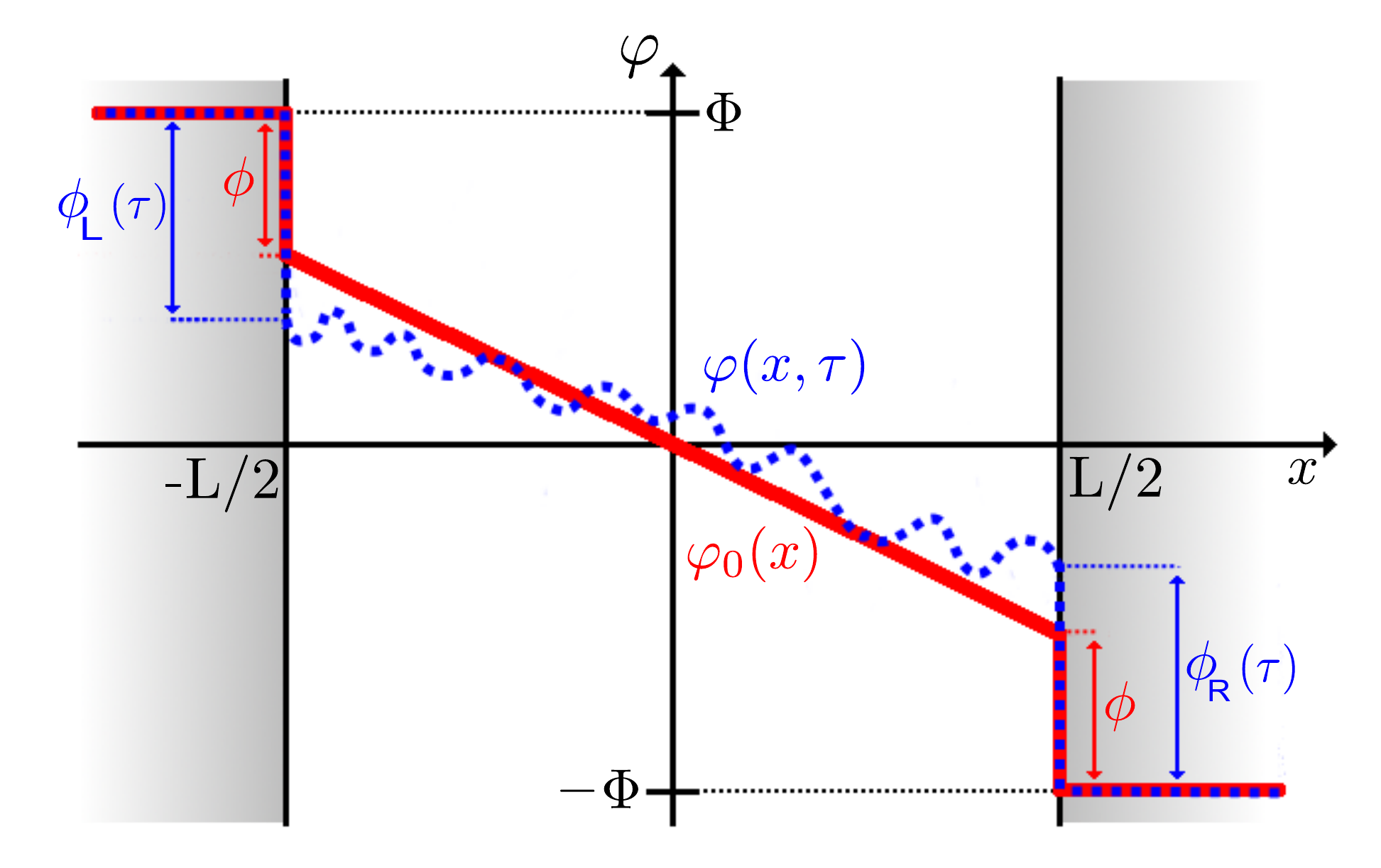}
\caption{(Color online) Phase profile along the
  channel. The solid line, $\varphi  _0({x})$ is a typical (symmetric) configuration made up of a linear superfluid contribution and phase
  jumps, $\phi  $, at  each tunnel barrier.  The dashed line represents a fluctuation around the phase profile.  Here we have chosen the phases of the BEC reservoirs as $\Phi _\mathrm{L}=-\Phi _\mathrm{R}\equiv \Phi   $. The phase profile shown above corresponds to  a phase difference of $2\Phi <\pi $.}\label{phasefig}
\end{figure}

\begin{figure}[t]
\centering
\includegraphics[width=0.95\columnwidth]{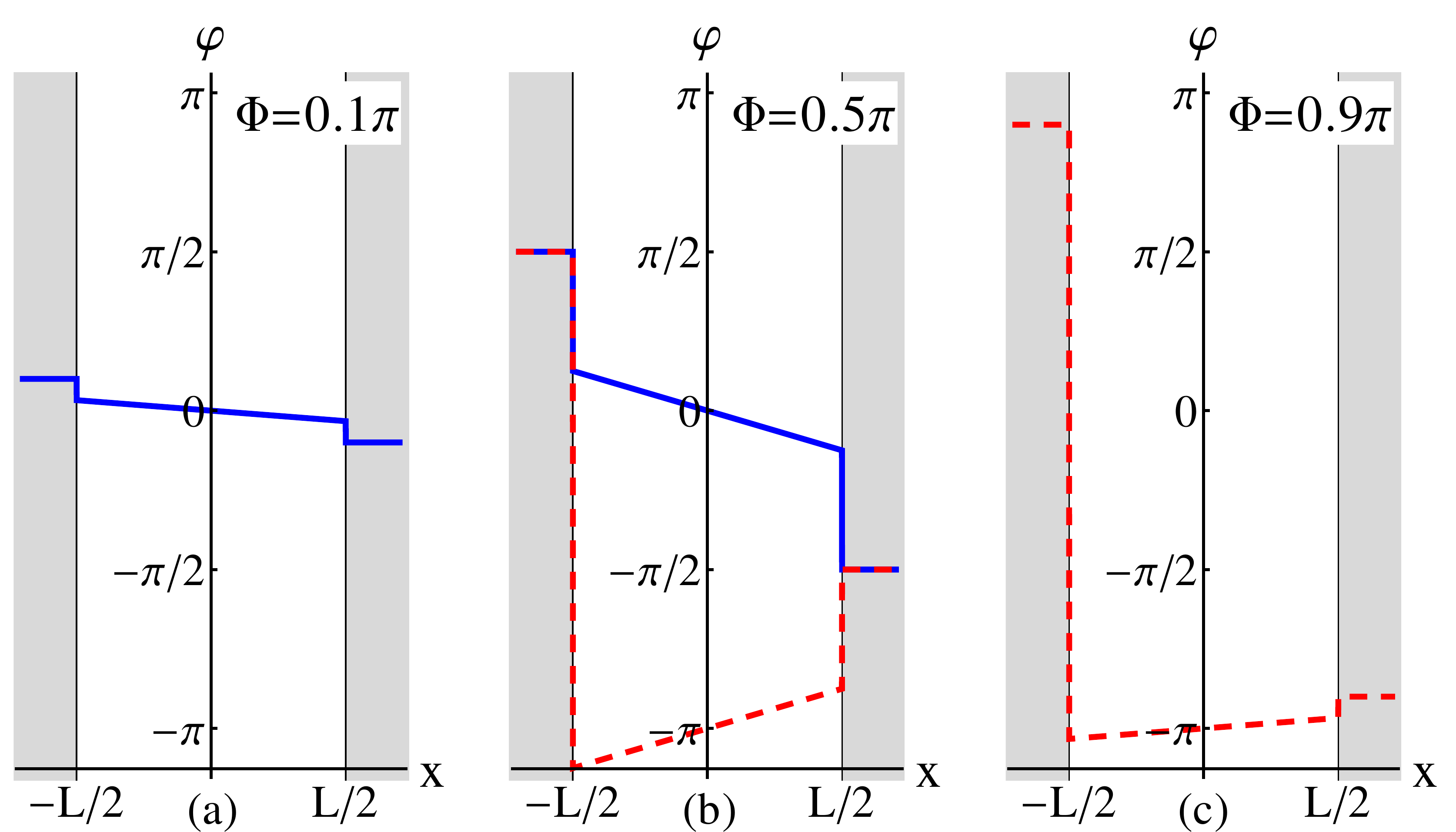}
    \caption{(Color online)  The MF phase profile in the channel for $\alpha\!<\!1$ ($J\!<\!J_\mathrm{c} $): (a) and (c) are unique symmetric/asymmetric solutions near $\Phi\!=\!0 $ or $\pi $, respectively; (b)~these two solutions become degenerate at $\Phi =\pi /2$,   with one of them  becoming metastable slightly above or below $\pi/2$.\label{MultiSol} }
\end{figure}

We demonstrate that the bosonic flow behaves drastically different to its condensed matter counterpart, i.e., an electronic flow between two bulk superconductors weakly connected by a 1D channel via Josephson junctions  \cite{Fazio1,*Fazio,*MasStoneGoldbartLoss,*AffleckCauxZ,Caux1,*Caux2}.
We show that  qualitatively new physics emerges here. The external phase difference between the reservoirs governs the phase profile illustrated in Fig.~\ref{phasefig}:  substantial phase drops at the tunneling barriers are followed by  a constant superflow of the quasicondensate through the 1D channel.  Such a superflow  is parametrically larger than that expected from a perturbative approach, which is appropriate for the corresponding electronic case \cite{Fazio} but totally fails for the bosonic superflow.
Surprisingly, for an external phase difference  close to $\pi $, the phase profile  turns out to be always bistable so that the superflow can spontaneously change direction (see Fig.~\ref{MultiSol}). With increasing the tunneling, such a bistability spreads to all values of $\Phi $. This would   lead to jumps and hysteresis in the sawtoothlike observable superflow, making it qualitatively different from an almost sinusoidal Josephson supercurrent in the corresponding superconducting systems.

The geometry sketched in Fig.~\ref{fig:1}, required for observing  these phenomena in flows of ultracold bosons, can be experimentally implemented by exploiting the versatility of potential shaping on atom chips \cite{Folman-Kruger-et-al,*FKS02}. Here we can form two bulk reservoirs weakly connected by a  1D  channel and imprint an arbitrary phase difference between them, while keeping the chemical potentials equal \cite{SGLK:1}.

This scenario is a starting point for experimental studies of  different regimes of the bosonic superflow that we investigate theoretically in this Letter. We will show how the results described above are obtained from a mean-field approach and prove it to be robust against fluctuations.

We consider a system comprising two bulk reservoirs, each containing a BEC, which are coupled via a 1D channel separated from the reservoirs by  weak
tunnelling barriers, see
Fig.~\ref{fig:1}.  The BEC in the left and right reservoirs is described
by order parameters $\Psi_\mathrm{L,R} = \sqrt{N_\mathrm{L,R}} \mathrm{e}
^{i\Phi_\mathrm{L,R}}$. Without loss of generality, we choose $\Phi _\mathrm{L}=-\Phi _\mathrm{R} \equiv \Phi    $. We assume that the reservoirs have been equilibrated to the same chemical potential and thus have equal particle densities, $n_\mathrm{L}=n_\mathrm{R}$, so that the current through the channel is driven only by the phase difference $2\Phi $.

The $N$ bosons in a 1D channel of length $L$ form a quasicondensate described by an order  parameter $\psi (x,t) =\sqrt{n+\rho(x,t)} \mathrm{e} ^{i\varphi(x,t)}$, where $\varphi $ is a  phase field and $\rho $ denotes density fluctuations around the mean density $n=N/L$.  As $\rho $ and $\phi  $ are canonically conjugate,  the  imaginary-time action describing phononlike low-energy excitations can be written via $\varphi  $ alone and, assuming that $L\gg \xi$, has the standard Luttinger-liquid form \cite{Giamarchi,*GogNersTsv}:
\begin{equation}
S_\mathrm{LL}=\frac{K}{2\pi c}\int_{0}^{\beta}\!\mathrm{d}\tau\!
\int_{-L/2}^{L/2}\!\mathrm{d}x
\left[(\partial_{\tau}\varphi)^2 + c^2(\partial_x \varphi)^2\right] . \label{lut}
\end{equation}
 Here $\xi\equiv 1/mc$ is the healing length \footnote{Here and elsewhere in the Letter we use units with $\hbar =1$.},
$c$ is the sound velocity,  $m$ is the bosonic mass,  $K \!\equiv\!  \pi n \xi $ is the Luttinger parameter:  $K \!\geqslant\! 1$ for bosons with a short-range repulsion,  with $K\!=\!1$ corresponding to the Tonks-Girardeau gas of hard-core bosons equivalent to the ideal Fermi gas \cite{Giamarchi}. For typical experimental  situations, $\xi $ is   much larger than the distance between bosons, so that  $K\!\gg\!1$.

We model the coupling of the reservoirs to the channel by a tunneling action, assuming for simplicity\footnote{Allowing for a difference in the tunneling  energies at both barriers leads insignificant changes in parameters without qualitatively affecting our results \cite{SGLK:1}.} the tunneling  energies at both barriers being equal to $J$:
\begin{align}
\label{eq:tunnel}
S_\mathrm{T}&=2J \int_0^\beta\!\mathrm{d}\tau  \big[ \cos\phi_\mathrm{R} +
  \cos\phi_\mathrm{L}\big],
\end{align}
where  $\phi _\mathrm{L,R} $ are the expected phase drops   at the barriers ({$x=\pm   L/2$}).  As usual,  the tunneling action is valid when the overlap of the wave functions across the barrier is small,  which imposes  the requirement $J\ll cK/\xi\equiv \pi nc$.

A second order in $J$ perturbational calculation of the bosonic supercurrent gives a result divergent at $T\to0$  for the values of $K$ pertinent to bosonic systems. So, unlike  superconducting systems  \cite{Fazio},  for which the perturbative approach is fully adequate,   a nonperturbative treatment is  required here.

We start our analysis with finding a nontrivial mean-field (MF) configuration for  the model ({\ref{lut}}) and ({\ref{eq:tunnel}}).   The phase field $\varphi ({x})$ in the channel is related  to the phase drops at the barriers by the boundary conditions:
\begin{align}\label{phi}
    \phi_\mathrm{L} &=\Phi -\varphi(- L/2)\,, &\phi _\mathrm{R}&=\Phi +\varphi  (L/2)\,.
\end{align}
 Then we minimize the action ({\ref{lut}})--({\ref{eq:tunnel}})  by {a stationary} solution satisfying the above boundary condition:
\begin{align}\label{condition}
   \varphi_0(x)&=-\phi _-  -2({\Phi -\phi _+ })\frac{x}{L}\,,  & \phi_\mathrm{\pm}  &\equiv\frac{1}{2}({\phi _{\mathrm{L}}\pm\phi _\mathrm{R} })\,.
\end{align}
It describes a constant superflow, $\mathcal{I}= nv $,  between
the reservoirs, with a velocity $v=-2({\Phi -\phi _+})/mL$.
The energy $E$ is the sum of  the supercurrent kinetic energy,  $\frac12{mN}v^2$, which arises from the Luttinger action Eq.~(\ref{lut}) on substituting ansatz ({\ref{condition}}),  and the Josephson energy,
$
  - 2J(\cos\phi _\mathrm{R}+\cos\phi _\mathrm{L})\,.
$
The total  dimensionless energy, $\varepsilon  \equiv E/J_\mathrm{c} $,  can be written via the phase drops $\phi _\pm$ as
\begin{align}\label{eq:energy}
   \varepsilon  &=  \phantom{-}2({\Phi -\phi_+  })^2 -4\alpha \cos \phi_+\cos\phi _-  \,, &\alpha& \equiv  J/J_\mathrm{c}
\end{align}
where $J_\mathrm{c}\equiv n/mL \ll \pi  nc$  so that $\alpha $ can vary from $0$ to values $\gg1$ within the region of applicability of the tunneling Hamiltonian, Eq.~(\ref{eq:tunnel}).

 All possible MF solutions are obtained by minimizing $\varepsilon  $ with respect to
  $\phi_+$ and $\phi _- $ at a fixed $\Phi $ which gives
\begin{subequations}\label{MF}\begin{align}
  \label{eq:phiphi}
 &\Phi -\phi _+ = \alpha \sin\phi_+\cos\phi _-\,,\\
   &\cos\phi _+\sin\phi _- =0\,.\label{phi-}
\end{align}
Since energy ({\ref{eq:energy}}) is a $2\pi $ periodic function of $\phi  _-$, we can restrict ourselves to  two solutions of Eq.~(\ref{phi-}), corresponding to  the symmetric phase drops, $\phi  _-=0$ so that $\phi  _\mathrm{R}=\phi  _\mathrm{L}  =\phi  _+$, and asymmetric ones, $\phi  _-=\pi $ so that  $\phi  _\mathrm{L}  =\phi  _++\pi $. Solutions corresponding to $\cos\phi  _+=0$ are always unstable (saddle points). For the  symmetric/asymmetric branch Eq.~({\ref{eq:phiphi}}) is  reduced to
   \begin{align}\label{pm}
   \Phi -\phi  _+=\pm \alpha \sin\phi  _+
\end{align}\end{subequations}
 The symmetric-branch equation is almost identical to  that emerging in a text-book analysis of a  superconducting quantum interference device (SQUID)  \cite{Tinkham}; however, its solution has a peculiar $4\pi $ periodicity. It is the coexistence of this solution with that for the asymmetric branch which restores the correct $2\pi $ periodicity.  Indeed, each of Eqs.~(\ref{pm}) has at least one stable solution in some interval of $\Phi $ and, remarkably,  these intervals always overlap.

The MF energy   is thus no longer a single-valued function of $\Phi $. Assuming first a singly connected geometry, when the external phase difference $2\Phi \in[{0,2\pi }] $,
we find for small $\Phi $ that the lowest energy solution of Eq.~(\ref{eq:phiphi}), which is $\phi _+\approx \Phi/(1+ \alpha) $, belongs to the symmetric branch. An elementary analysis shows that for small $\alpha $ it remains stable with increasing $\Phi $  up to $\Phi =\pi /2+\alpha $. The lowest-energy solution around $\Phi =\pi $ belongs to the asymmetric branch and remains stable down to $\Phi =\pi /2-\alpha $. Thus, in the interval of width $2\alpha $ centred at \mbox{$\Phi =\pi /2$}  the two solutions   coexist: the symmetric solution is stable and the asymmetric is metastable at \mbox{$\Phi <\pi /2$}, with their roles reversing at $\Phi >\pi /2$, as illustrated in Fig.~{\ref{MultiSol}}.

\begin{figure}[t]
\centering
\includegraphics[width=0.95\columnwidth]{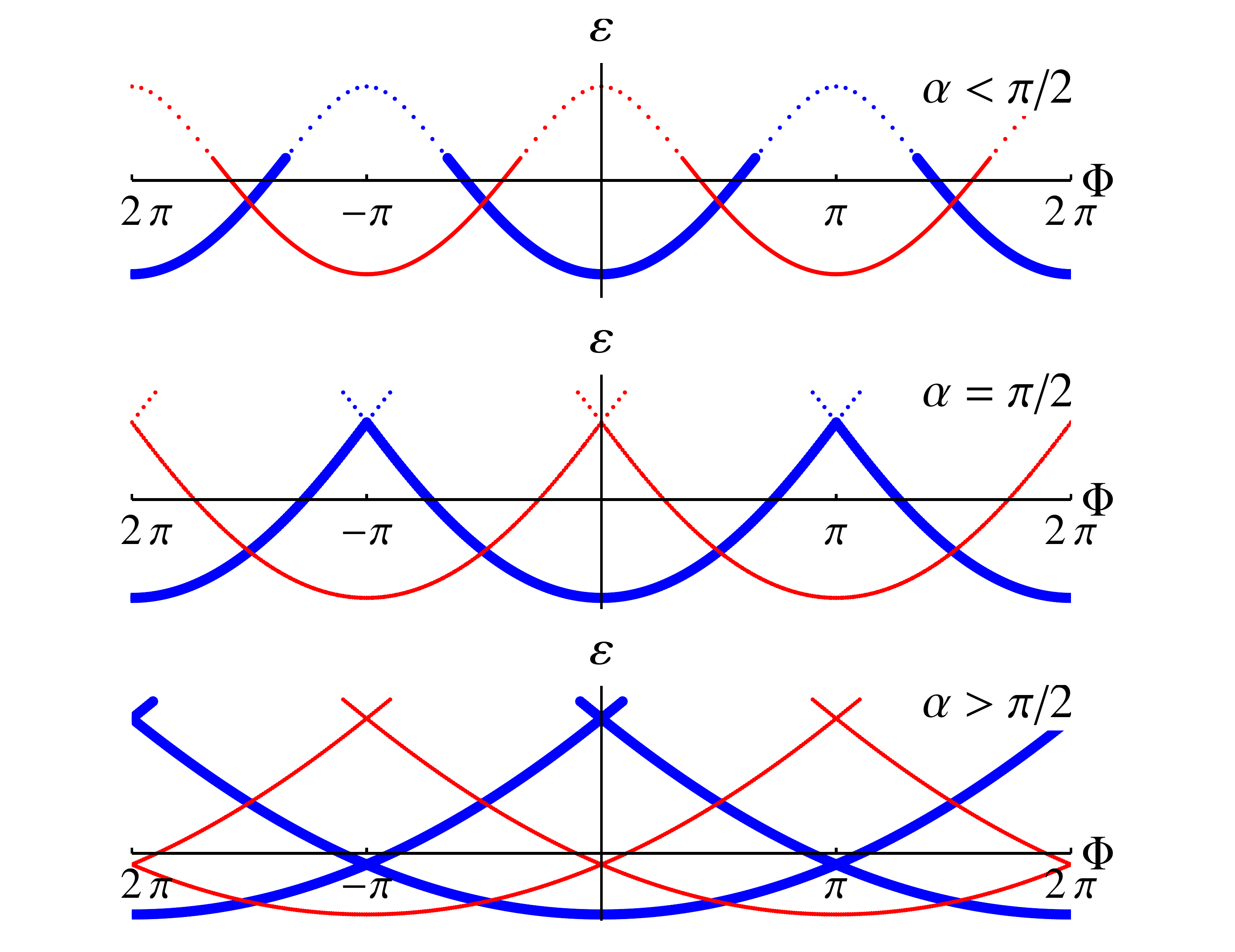}
(a)

\vspace*{12pt}

\includegraphics[width=0.95\columnwidth]{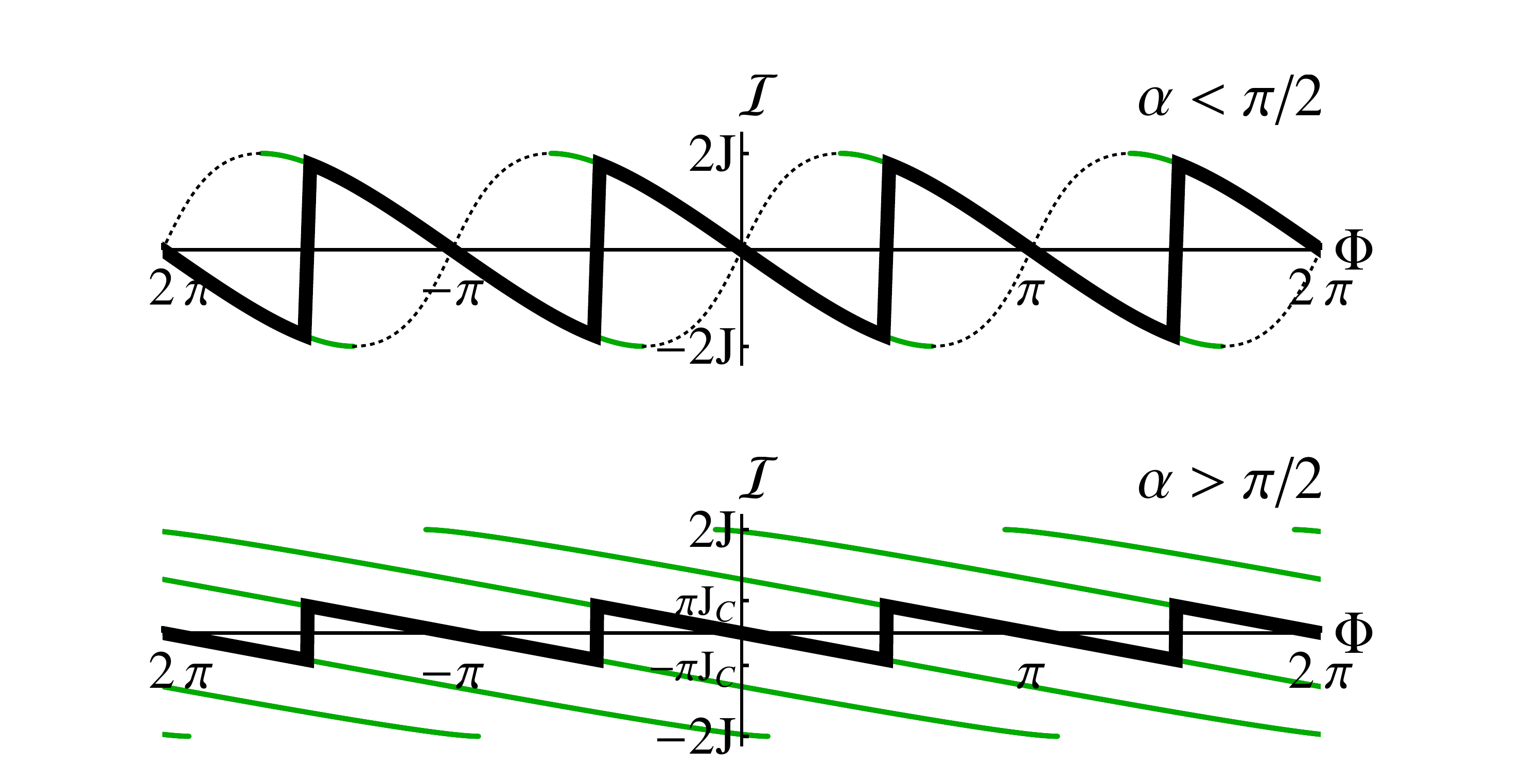}
(b)

\caption{(Color online) (a) The MF energies ({\ref{eq:energy}}), spread between $\varepsilon  _\mathrm{min}= -4\alpha $ and $ \varepsilon_{\mathrm{max} }=2\alpha ^2$,  as functions of the external phase difference, $2\Phi $, at different values of $\alpha$. Thick (thin) solid lines represent symmetric (asymmetric) stable or metastable   solutions,   the latter lying in the continuum of phononic excitations. (b) The dependence of the superflow on $2\Phi $, with thick (thin) lines representing stable (metastable) flow.  Dashed lines represent unstable solutions drawn  as a guide to the eye.
}\label{cusps}
\end{figure}

With $\alpha $ increasing, two new solutions appear at $\alpha >1$ for both the symmetric and asymmetric branch but they remain unstable until $ \alpha $ reaches $\pi /2$. At this point the two solutions coexist in the entire   interval $[{0,\pi }] $, while new metastable solutions emerge for the asymmetric branch around $\Phi =0$ and for the symmetric around $\Phi =\pi $. With $\alpha $ further increasing, new pairs of metastable solutions appear at integer multiples of $\pi /2$, see Fig.~{\ref{cusps}}.

 It follows from Eqs.~(\ref{condition}) and ({\ref{eq:phiphi}}) that the  superflow along the channel is $\mathcal{I}= \mp2J \sin\phi_+   $. As the sign comes from $\cos \phi  _- =\pm1$, it is easy to see that this corresponds to the sum of the Josephson currents across the barriers, $-J({\sin \phi  _\mathrm{L} +\sin \phi  _\mathrm{R} })$, as expected. What is non-trivial is the relation of this to the external phase difference, $2\Phi $, given by Eqs.~(\ref{MF}).

 For small $\alpha $ (i.e.\ for $J\ll J_\mathrm{c} $),  $\phi  _+\approx \Phi\mp\alpha \sin \Phi  $ so that almost the entire phase change accumulates  at the Josephson barriers. The phase drops look very different for the two branches, symmetric with $\phi _-=0$ and asymmetric with $\phi _-=\pi$. In the former case,   $\phi  _{\mathrm{L} }=\phi  _{\mathrm{R} }$ by definition   while in the latter $\phi  _{\mathrm{L} }=\Phi +\pi  $ and $\phi  _\mathrm{R}=\Phi -\pi  $. This means that, e.g., near the energy minimum $\Phi =\pi $, almost the entire phase drop, $2\pi $, occurs at  one of the barriers. The phase profiles described by these two branches correspond to the superflows $\mp 2J\sin \Phi  $, each  being  $4\pi $ periodic  with respect to the overall phase difference $2\Phi $. As the symmetric branch is stable for $2\Phi <\pi +2\alpha $ and asymmetric for $2\Phi >\pi -2\alpha $, the correct $2\pi $ periodicity is restored by jumps between the branches which can occur anywhere in the intervals of coexistence.

With  $\alpha $ increasing, the metastable energy solutions are reflected in the superflow, $\mathcal{I}=-J_\mathrm{c} \dd \varepsilon  /\dd (2\Phi )  $, Fig.~\ref{cusps}(b). The superflow corresponding to the lowest energy configuration changes from the piecewise sinusoid for $\alpha <\pi /2$ to a sawtooth function at $\alpha >\pi /2$, given by $\mathcal{I}=-2J_\mathrm{c} \Phi   $ for $\Phi \in[{-\pi /2,\pi /2}] $ and periodically repeated for all $\Phi $. In the latter case, when $J\gg J_\mathrm{c} $, the maximal possible superflow saturates at $\mathcal{I} =\pi J_\mathrm{c} $.
Such a characteristic saw-tooth shape for any value of the tunneling is an inevitable consequence of the metastability.  In contrast, for the case of superconductors connected by a LL channel via two JJ (corresponding in our notations to $K\!<\!1/2$), the perturbative Josephson current has a  slightly distorted sinusoidal shape \cite{Fazio1}. It is interesting that the exact solution for the boundary case, {$K\!=\!1/2$}, shows a crossover from a smooth to a saw-tooth shape with increasing the tunneling \cite{Caux1}.

The existence of metastable solutions should reveal itself experimentally in  hysteresis of the superflow, as we will discuss at the end of the Letter.

It is important that phonon fluctuations in the channel do not wash out essential features of the MF solutions, Eqs.~({\ref{condition}})--(\ref{MF}), and remarkable that they do not result in avoided crossings in Fig.~\ref{cusps}(a). To show this we introduce the phase fluctuations in the 1D channel, $\widetilde\varphi(x,\tau)= \varphi(x,\tau)-\varphi_0(x)$ and at the boundaries, $\widetilde\phi  _{\mathrm{L,R} }(\tau)= \phi  _{\mathrm{L,R} }(\tau)-\phi_{\mathrm{L,R} }$, related by the boundary conditions $\widetilde\varphi(\pm L/2,\tau) = \pm \dphi_{\mathrm{L,R} }(\tau)$. Here
$\varphi  _0({x})$ and $\phi_{\mathrm{L,R} } $ are the  solutions of the MF equations ({\ref{MF}}) described above, related to the symmetric and antisymmetric combinations  introduced in Eq.~(\ref{condition}). Then, after integrating out the  Gaussian fluctuations in the 1D channel, we obtain the effective action, $S=S_\mathrm{fl}+S_\varepsilon   $:
\begin{subequations}\label{effectiveaction}
\begin{align}\label{Ohm}
 S_\mathrm{fl}
 &= K\!\int \! \frac{|\omega |\mathrm{d}\omega}{2\pi^2} \left[    |\dphi _+(\omega)|^2 +    |\dphi_  -(\omega)|^2 \right],
 \\\label{eps}
  S_\varepsilon  &
 =\!\int \!\! \mathrm{d}\tau  \,\varepsilon[{\phi  _+({\tau }),\phi  _-({\tau }); \Phi  }] .
\end{align}
\end{subequations}
Here $\varepsilon  $ is the function of  $\phi  _{\pm }({\tau })=\dphi_\pm({{\tau }})+\phi  _\pm   $ and is given by Eq.~(\ref{eq:energy}). It plays the role of an effective ``washboard'' potential for the Caldeira-Legget type   action of Eq.~(\ref{Ohm}).  We assumed in deriving Eq.~(\ref{Ohm}) that $\omega \gg \pi c/L$, which is the lowest phonon energy in the channel \cite{SGLK:1}.

Now we perform the standard renormalization group  (RG) analysis by integrating out fast modes in the fields $\phi  _+ $ and $\phi  _- $, as described for completeness in Supplemental Material \cite{SGLK:1}. This  results in the RG equation for the dimensionless tunneling strength $\alpha   $:
\begin{align}\label{RG}
\frac{\dd \ln\alpha }{\dd \ln b}=1- \frac{1}{2K},
\end{align}
where b is a scaling parameter.  The integration between the upper, $\Lambda \sim c/\xi $, and lower, {$\omega _0\sim \max\{{T,\, c/L}\}  $}, energy cutoffs  gives the renormalized dimensionless tunneling as $\alpha ({\omega _0 }) = \alpha _0 \left( \Lambda /\omega _0 \right)^{1-\frac{1}{2K}}$, where $\alpha _0\equiv J/J_\mathrm{c} $. Since the tunneling through barriers separated by $L\gg\xi$ is uncorrelated, this is similar to the results   for  tunneling through a single barrier \cite{Kane-Fisher}, as well as to the results for superconducting systems \cite{Fazio} in a geometry similar to that under consideration here.

A remarkable feature is that for $K\gg1$, characteristic of ultracold bosonic systems with the healing length much bigger than the interatomic distance,  $\alpha $ flows to larger values. This means that the washboard potential  becomes more pronounced so that the fluctuations are  irrelevant in the low-energy limit and the MF solution, described above, is robust. In particular, since the fluctuations do not connect different MF branches, the level crossings are not avoided and the characteristic cusps in energy, Fig.~\ref{cusps}, and the corresponding jumps in the superflow remain.  Alternatively, this can be seen using instanton techniques similar to those of Ref.~\cite{Buchler2001,*Schecter2012}. Namely, the probability of an instanton connecting two degenerate configurations, like in  Fig.~\ref{MultiSol}(b), can be shown to be vanishingly small for $K\gg1$.

 Experimental data about the superflow can be extracted from images taken of the atomic density distribution in the channel at different times throughout the evolution of the system. The phase imprinting can be implemented in two different ways. First, we can imprint the phase difference before connecting the reservoirs thus mapping the lowest, stable branches of the energy (Fig.~\ref{cusps}), and measuring jumps in the superflow direction. Secondly, we can gradually modify the phase difference \emph{in vivo}, with the weak link already present, thus being able to explore the metastable branches by observing a hysteretic behavior in the superflow.

Complementary direct measurements of the phase profile are possible  by keeping part of the bulk BEC as a homogenous phase reference.  Then a readout can be obtained from  an interference pattern between this reference and the  quasicondensate in the channel \cite{SGLK:1}.

In conclusion, we have demonstrated that bosonic superflow driven by a phase difference between two BEC reservoirs  has spectacular features without any analogy in geometrically similar superconducting systems. The superflow, which is proportional to the first (rather than second) power of the tunneling energy, periodically flips direction and, moreover, has metastable branches, Fig.~\ref{cusps}. The corresponding energy   levels intersect, and fluctuations do not lead to avoided crossings. The bi- and multistability associated with the  existence of metastable branches can only be accessed  dynamically. A theoretical description of the kinetics of such a process, while going beyond the scope of this Letter, remains ad interesting open question. Experimentally, the multistability can be revealed by  gradually adjusting the phase difference between the reservoirs at finite tunneling.

\begin{acknowledgments}
We gratefully acknowledge support from  the Leverhulme Trust via the Grant No.\ RPG-380 (I.V.L.) and from the EPSRC.
\end{acknowledgments}

 \advance \textheight by -6cm

\newpage\advance\textheight by 6cm

\begin{widetext}
    \begin{appendix}
\section*{Supplemental Online Material}

\section{Fluctuational Action}
We consider fluctuations around the mean field (MF) solution, $\varphi_0  ({x})$, given by Eqs.~(\ref{condition})-(\ref{MF}) of the main text.  The fluctuations in the one dimensional (1D) channel are defined as $\widetilde{\varphi} (x,\tau) = \varphi(x,\tau) - \varphi_0 (x)$ and the independent fluctuations at the boundaries are $\widetilde{\phi}_{L,R}(\tau) = \phi_{\mathrm{L,R} }(\tau) -\phi_{\mathrm{L,R} }$.
These fluctuations must obey the same boundary conditions as the MF solution so that $\widetilde{\varphi} (-L/2,\tau ) = - \widetilde{\phi}_L(\tau)$ and $\widetilde{\varphi} (L/2,\tau ) =  \widetilde{\phi}_R(\tau)$.  It is convenient for what follows to introduce the field combinations $2\widetilde{\phi}_{\pm} (\tau) = \widetilde{\phi}_L (\tau) \pm \widetilde{\phi}_R (\tau)$, as in Eq.~(\ref{condition}) of the main text.

It is simple to see that the tunneling action, Eq.~(\ref{eq:tunnel}), is given in terms of these fluctuations as:
\begin{align}\label{1}
S_{J} &= 2J \int \!\mathrm{d}\tau \, \left[ \cos\phi_L(\tau) + \cos\phi_R(\tau) \right]= 4J \int \!\mathrm{d}\tau \, \cos(\phi_{+} + \widetilde{\phi}_{+}(\tau)) \cos(\phi_{-}+\widetilde{\phi}_{-}(\tau)).
\end{align}
We now consider the fluctuations   in the 1D channel.  After taking into account the   boundary conditions, the phase field in the 1D channel is $\varphi(x,\tau) = -\phi_{-}(\tau) - \frac{2x}{L}\left( \Phi - \phi_{+}(\tau) \right) + \sum_{n=1}^{\infty} \left[ \varphi_n^{\mathrm e}(\tau) \cos \frac{(2n-1)\pi x}{L} + \varphi_n^{\mathrm o}(\tau)\sin\frac{2 \pi n x}{L} \right] $.  Here the first two terms are the fluctuating counterparts of the MF solution while the remainder, which satisfies the Dirichlet boundary conditions at $x=\pm L/2$, is Fourier expanded. Substituting this into the Luttinger action  (\ref{lut}) and performing the Fourier transform in $\tau $ gives $S = S_{\varepsilon} + S_{0} + S_{\mathrm{o} } + S_{\mathrm{e} }$, where
\begin{align}\label{S-eps}
S_{\varepsilon} &= \int \mathrm{d}\tau \left[ 2J_\mathrm{c}(\Phi-\phi_{+})^2 - 4J_\mathrm{c}(\Phi-\phi_{+})\widetilde{\phi}_{+}(\tau)  + 4J\cos(\phi_{+} + \widetilde{\phi}_{+}(\tau)) \cos(\phi_{-}+\widetilde{\phi}_{-}(\tau)) \right]
\end{align}and\begin{align}
S_{0 } &= \frac{Kc}{2\pi}\int \frac{\mathrm{d}\omega}{2\pi} \Bigg[ \left(\frac{4}{L}+ \frac{L\omega^2}{3c^2}\right) \bigl|\widetilde{\phi}_{+}(\omega)\bigr|^2 + \frac{L\omega^2}{c^2}\bigl|\widetilde{\phi}_{-}(\omega)\bigr|^2  \Bigg]
\nonumber \\
S_{\mathrm{o} } &=\frac{Kc}{2\pi}\int \frac{\mathrm{d}\omega}{2\pi} \sum_{n=1}^{\infty}  \left[ \frac{1}{2}\left( \frac{4\pi^2 n^2}{L} + \frac{L\omega^2}{c^2} \right)\bigl|\varphi_n^{\mathrm o}(\omega)\bigr|^2  - \frac{2L(-1)^n \omega^2}{n\pi c^2} \widetilde{\phi}_{+}(\omega) \varphi_n^{\mathrm o}(-\omega)\right]\label{S2}  \\
S_{\mathrm{e} } &= \frac{Kc}{2\pi}\int \frac{\mathrm{d}\omega}{2\pi}\sum_{n=1}^{\infty} \left[ \frac{1}{2}\left( \frac{\ (2n-1)^2 \pi^2}{L} + \frac{L\omega^2}{c^2} \right)\bigl|\varphi_n^{\mathrm e}(\omega)\bigr|^2  + \frac{4L(-1)^n \omega^2}{\pi(2n-1) c^2} \widetilde{\phi}_{-}(\omega) \varphi_n^{\mathrm e}(-\omega)\right].\nonumber
\end{align}
The only non-Gaussian part of the action, Eq.~(\ref{S-eps}), does not depend on the fields in the channel, $\varphi  ^\mathrm{e} $ and $\varphi  ^\mathrm{o} $, so that they can be integrated out. By symmetry, the even and odd parts of the fluctuational field, $\varphi  ^\mathrm{e} $ and $\varphi  ^\mathrm{o} $, are not mixed and can be  integrated out independently.  Integrating out the odd fluctuations gives
\begin{align}
\int \! \mathcal{D}\varphi_n^{\mathrm o} \, e^{-S_{\mathrm{o} }[\varphi_n^{\mathrm o}]} &= \exp \left[ \frac{K }{2\pi c} \int \! \frac{\mathrm{d}\omega}{2\pi} \, \sum_{n=1}^{\infty} \frac{2 L^3 \omega^4 |\widetilde{\phi}_{+}(\omega)|^2}{ n^2 \pi^2  \left(  {4\pi^2 n^2c^2}  +  {L^2 \omega^2}  \right) } \right]  \nonumber \\
& = \exp \left[-\frac{K }{2\pi  } \int \!  \frac{\mathrm{d}\omega}{2\pi} \,|\widetilde{\phi}_{+}(\omega)|^2 \left( {2\omega} \coth\left(\frac{\omega L}{2 c}\right) - \frac{4c}{L} - \frac{\omega^2 L}{3 c } \right)\right].\label{So}
\end{align}
Similarly, integrating out the even fluctuations gives
\begin{align}
\int \! \mathcal{D}\varphi_n^{\mathrm e} \, e^{-S_{\mathrm{e} }[\varphi_n^{\mathrm e}]} &= \exp \left[ \frac{K }{2\pi c} \int \! \frac{\mathrm{d}\omega}{2\pi} \, \sum_{n=1}^{\infty} \frac{8 L^3 \omega^4 |\widetilde{\phi}_{-}(\omega)|^2}{  (2n-1)^2 \pi^2   \left[  {(2n-1)^2\pi^2 }{c^2} + {L^2 \omega^2}  \right] } \right]  \nonumber \\
& = \exp \left[-\frac{K }{2\pi} \int \!  \frac{\mathrm{d}\omega}{2\pi} \,|\widetilde{\phi}_{-}(\omega)|^2 \left( {2\omega} \tanh\left(\frac{\omega L}{2 c}\right) - \frac{\omega^2L }{c } \right)\right].\label{Se}
\end{align}
Combining Eqs.~(\ref{So}) and ({\ref{Se}}) with action $S_0 $, i.e.\ the first line of Eq.~(\ref{S2}), gives the full  fluctuational action in terms of the fluctuating boundary fields:
\begin{align}
S_\mathrm{fl} &= \int \! \frac{\mathrm{d}\omega}{2\pi} \, \frac{K \omega}{\pi} \left[ \coth\left(\frac{\omega L}{2 c}\right) \bigl|\widetilde{\phi}_{+}(\omega)\bigr|^2  + \tanh\left(\frac{\omega L}{2 c}\right) \bigl|\widetilde{\phi}_{-}(\omega)\bigr|^2 \right]
\end{align}
This action can be further simplified at relevant energies, $\omega \gg c/L \equiv  \omega_0$ which is the lowest phononic energy in the system.  The fluctuational action is then
\begin{equation}\label{S-fl}
S_\mathrm{fl} = \frac{K}{\pi} \int \! \frac{\mathrm{d}\omega}{2\pi} |\omega| \left[ |\widetilde{\phi}_{+}(\omega)|^2 + |\widetilde{\phi}_{-}(\omega)|^2 \right]\equiv  \frac{K}{2\pi} \int \! \frac{\mathrm{d}\omega}{2\pi} |\omega| \left[ |\widetilde{\phi}_{\mathrm L}(\omega)|^2 + |\widetilde{\phi}_{\mathrm R}(\omega)|^2 \right].
\end{equation}
in accordance with Eq.~(\ref{Ohm}) of the main text. The full action (\ref{effectiveaction}) will be used for an RG analysis with $\omega _0$ playing the role of the infrared cutoff and $\mu \equiv c/\xi $ the ultraviolet cutoff.

\section{RG Analysis}
We perform the standard renormalization group (RG) analysis of the fluctuational action  Eq.~(\ref{effectiveaction}), which is equal to the sum of actions ({\ref{S-eps}}) and ({\ref{S-fl}}). It is convenient to do this in terms of the original fields $\phi  _{\mathrm{L,R} }({\tau }) = {\phi  _+({\tau })\pm\phi  _-  }({\tau })  $, describing phase drops on the left and right barrier. To this end, we split the fields into the  fast and slow modes, $  \phi  _\mathrm{L,R}({\tau })=  \phi  _\mathrm{L,R}^>({\tau })+  \phi  _\mathrm{L,R}^< ({\tau })$,  comprising the Fourier components with energies $\Lambda/b < |\omega| < \Lambda$ and  $ |\omega|<\Lambda/b$, respectively.  Then we  average the non-Gaussian part of the action, Eq.~(\ref{1}), over the fast fluctuations, i.e.\ using $\langle{\ldots}\rangle_{ >}\equiv \int\mathcal{D}\phi  ^>({\ldots})e^{-S_\mathrm{fl}^> } /\int\mathcal{D}\phi  ^> e^{-S_\mathrm{fl}^> }  $, where $S_\mathrm{fl}^>$ is the fast part of action ({\ref{S-fl}}). Applying the identity $\langle{e^{i\phi  ^>}}\rangle_> =\exp[-\frac{1}{2}\langle{\phi  ^>}\rangle^2  ]=b^{-1/2K}$, we find the non-Gaussian part of the action is renormalized to first order in $J$ as follows:
\begin{align}\label{rg}
    2J\int\mathrm{d}\tau  \left\langle{\cos[{\phi  _\mathrm{L,R}^<({\tau })+{\phi  _\mathrm{L,R}^>}({\tau })}]}\right\rangle _> = b^{1-\frac{1}{2K} }2J\int \mathrm{d}\tau  \cos \phi  _\mathrm{L,R}^<({\tau })\,.
\end{align}
Differentiating this with respect to $\ln b$, we obtained the RG equation ({\ref{RG}}) given in the main text. Note that since the parts of the action corresponding to the phase drops on the left and on the right are renormalized independently, introducing different tunneling energies for the two barriers will not affect our conclusions.  This can also be seen by considering the MF solution in the presence of asymmetric tunneling barriers, $\alpha_{\mathrm{L}}\neq \alpha_{\mathrm{R}}$.  Following the procedure outlined in the main text, the equations minimizing the asymmetric MF energy are
\begin{subequations}\label{asymMF}\begin{align}
 &\frac{\Phi_{\mathrm{L}} - \Phi_{\mathrm{R}}}{2} -\phi _+ = \frac{2 \alpha_{\mathrm{L}}\alpha_{\mathrm{R}}}{\alpha_{\mathrm{L}}+\alpha_{\mathrm{R}}} \sin\phi_+\cos\phi _-\, \label{eq:asymMFa},\\
   &\cos\phi _+\sin\phi _- = \frac{\alpha_{\mathrm{L}} - \alpha_{\mathrm{R}}}{\alpha_{\mathrm{L}} + \alpha_{\mathrm{R}}}\sin\phi_+\cos\phi _- \label{eq:asymMFb} \,,
\end{align}
\end{subequations}
analagous to Eqs.~(\ref{MF}) for the symmetric tunneling.  A direct comparison of these equations reveals that the symmetric $\alpha$ of Eq.~(\ref{eq:phiphi}) is simply replaced by the harmonic average of the asymmetric $\alpha_{\mathrm{L}}$ and $\alpha_{\mathrm{R}}$ in Eq.~(\ref{eq:asymMFa}).  It can also be seen that Eq.~(\ref{eq:asymMFb}) defines two solutions on $[0,2\pi)$ which are shifted away from the  solutions, $\phi_{\mathrm{a}} = 0,\pi$ given in Eq.~(\ref{phi-}) for $J_\mathrm{R}=J_\mathrm{L}$, but remain exactly separated by $\pi$ so that $\phi_{\mathrm{a}}$ always acts as a label for solutions with $\pm \alpha_{\mathrm{L(R)}}$ in Eq.~(\ref{eq:asymMFa}).  Thus, we see that all essential features of the MF solution described in the main text are retained for the case of asymmetric tunneling, with only a simple change of parameters.

\section{Experimental implementation}

The suggested experimental implementation is based on an atom chip, where surface-mounted microfabricated current carrying wires can be employed to form a wide variety of trapping and guiding potentials. The starting point is a moderately elongated initial reservoir trap, in which a BEC of (essentially) three-dimensional nature will be formed. Two parallel Z-shaped wires carrying copropagating currents provide the necessary inhomogeneous fields. Unlike the standard case, the width of the central section of these wires is chosen to vary as a function of position $x$ along the trap as indicated in Fig. \ref{fig:sup_exp_1}. The size of the currents together with the strength of an external homogeneous bias field $B$ with a direction parallel to the surface plane determines the surface-trap separation $H$.

\begin{figure}[t]
\centering
a) \includegraphics[width=0.475\columnwidth]{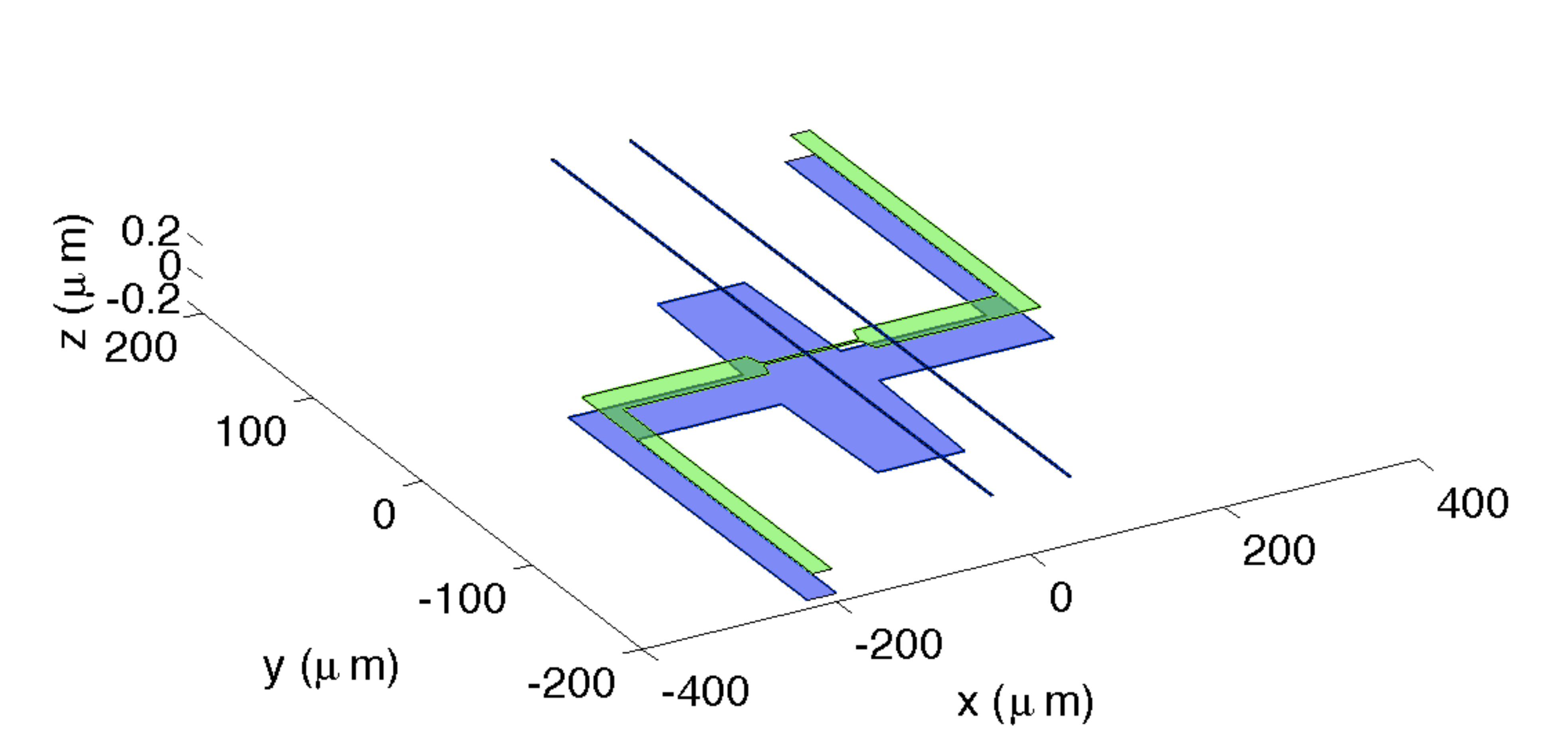}
b) \includegraphics[width=0.475\columnwidth]{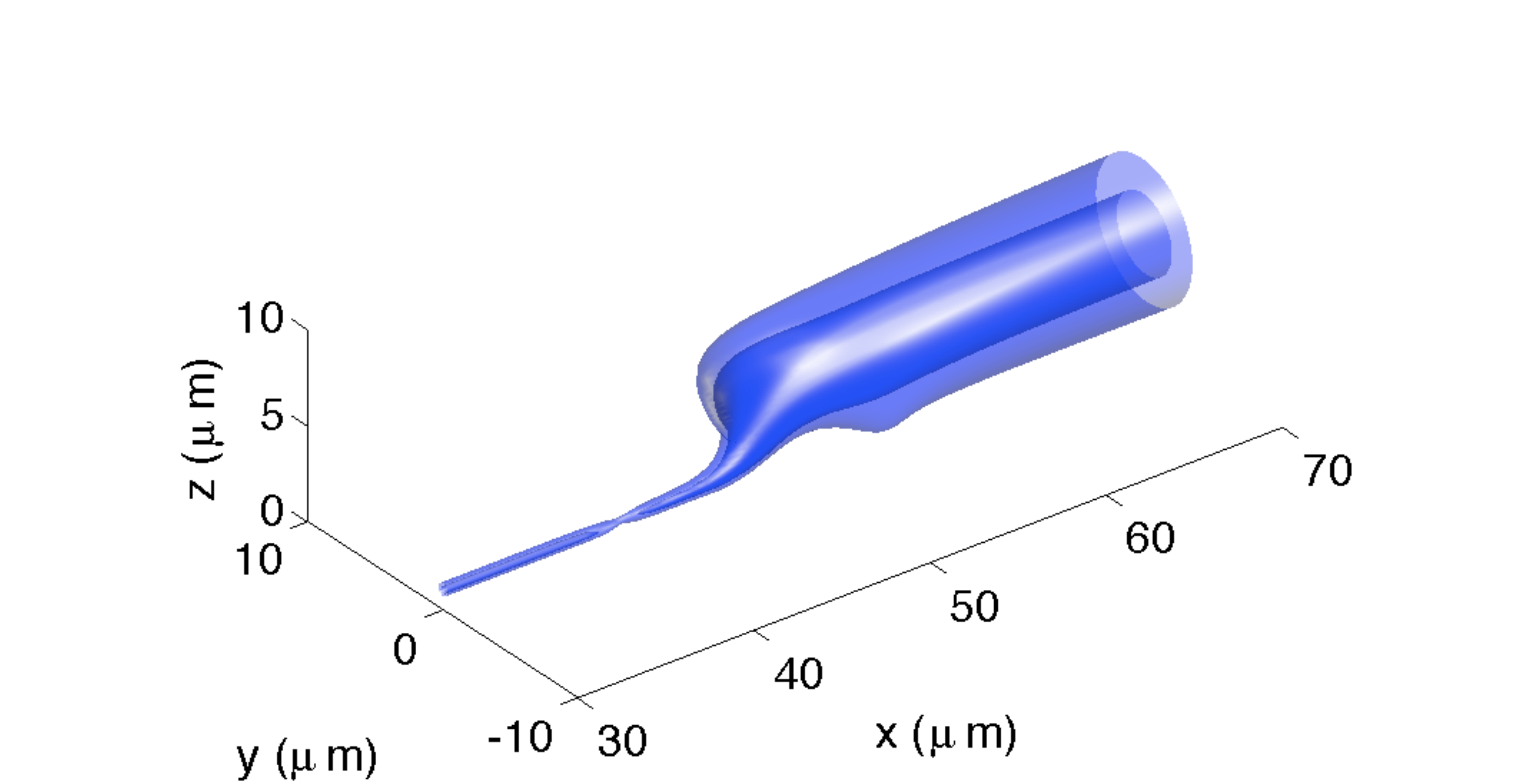}
\caption{a) Wire layout in three vertically separated layers on an atom chip. The lower two layers contain Z-shaped wires whose width varies as a function of position as shown. The top layer features two thin wires carrying an independent current that forms barriers of controllable height. b) Typical equipotential surfaces of the trap section where the 3d reservoir is connected to the 1d channel including one of the barriers. Here, the wire currents and external bias field are adjusted to trap the atoms at surface distances of a few micrometers. In contrast, at larger distances of tens to 100 microns a single 3d trap is formed in the same geometry.}\label{fig:sup_exp_1}
\end{figure}

For initial loading, $H$ is chosen larger than all wire widths and the distance between wires, so that a single simple 3d reservoir trap is formed. Subsequent adiabatic increase in $B$, decrease in the wire currents and corresponding reduction of $H$ will transfer the cloud into two equal 3d reservoirs that are connected through a narrow 1d channel. Additional adjustable currents through thin wires in the direction orthogonal to the channel positioned near its two connection points to the reservoirs are then used to introduce tunable barriers. Raising these barriers and then temporarily introducing a field gradient along $x$ slightly imbalancing the reservoirs will imprint a differential phase between the subclouds trapped in each reservoir. The raised barriers will prevent a chemical potential imbalance between the reservoirs during this preparation procedure.

Information about the suppercurrent, $\mathcal{I}=\rm dE/\rm d\Phi$ can be derived from images taken of the atomic density distribution in the channel at different times throughout the evolution of the system. Absorption images taken after several ms time-of-flight after all confining potentials have been switched off have been demonstrated to reach a sensitivity on the order of 3 atoms/$\mu$m \cite{Smith11}. Due to the strong transverse confinement in the channel, the line density as determined from the integral over the transverse dimensions is essentially unaffected by the time-of-flight. Fluorescence imaging through a sheet of near-resonant light spanned a few mm below the atom chip reaches even single atom sensitivity in the low-density regime \cite{Plisson} .  The presence of the cusps would be indicated by jumps in the supercurrent which would provide experimental support for our mean field solution.

As scheme to measure the phase profile predicted in the main text we envision to extend the setup discussed above to an interference experiment. After initial loading of a single reservoir trap, radio-frequency dressing the magnetic trapping potential can be used to vertically (along $z$) split the cloud into two \cite{Schumm2005}. One cloud is moved close to the surface where the shape of the trapping wires induces the formation of two reservoirs connected by a narrow channel in $x$-direction. The other cloud is moved away from the surface, so that it maintains its 3d BEC character with a homogeneous reference phase. At sufficiently large $z$-splitting distances, coherence between the clouds will not be maintained, so that the distant BEC provides an independent phase reference. For readout, both clouds are released from the trapping potential, so that they expand (essentially only in the $yz$-plane) and overlap. An interference pattern, again detected in absorption or, more sensitively, in fluorescence imaging, will form with a random global phase. However, the phase pattern along $x$ in the transport channel will be revealed in an inhomogeneous relative phase pattern along $x$.

\end{appendix}

\end{widetext}

\begin{thebibliography}{47}%
\makeatletter
\providecommand \@ifxundefined [1]{%
 \@ifx{#1\undefined}
}%
\providecommand \@ifnum [1]{%
 \ifnum #1\expandafter \@firstoftwo
 \else \expandafter \@secondoftwo
 \fi
}%
\providecommand \@ifx [1]{%
 \ifx #1\expandafter \@firstoftwo
 \else \expandafter \@secondoftwo
 \fi
}%
\providecommand \natexlab [1]{#1}%
\providecommand \enquote  [1]{``#1''}%
\providecommand \bibnamefont  [1]{#1}%
\providecommand \bibfnamefont [1]{#1}%
\providecommand \citenamefont [1]{#1}%
\providecommand \href@noop [0]{\@secondoftwo}%
\providecommand \href [0]{\begingroup \@sanitize@url \@href}%
\providecommand \@href[1]{\@@startlink{#1}\@@href}%
\providecommand \@@href[1]{\endgroup#1\@@endlink}%
\providecommand \@sanitize@url [0]{\catcode `\\12\catcode `\$12\catcode
  `\&12\catcode `\#12\catcode `\^12\catcode `\_12\catcode `\%12\relax}%
\providecommand \@@startlink[1]{}%
\providecommand \@@endlink[0]{}%
\providecommand \url  [0]{\begingroup\@sanitize@url \@url }%
\providecommand \@url [1]{\endgroup\@href {#1}{\urlprefix }}%
\providecommand \urlprefix  [0]{URL }%
\providecommand \Eprint [0]{\href }%
\providecommand \doibase [0]{http://dx.doi.org/}%
\providecommand \selectlanguage [0]{\@gobble}%
\providecommand \bibinfo  [0]{\@secondoftwo}%
\providecommand \bibfield  [0]{\@secondoftwo}%
\providecommand \translation [1]{[#1]}%
\providecommand \BibitemOpen [0]{}%
\providecommand \bibitemStop [0]{}%
\providecommand \bibitemNoStop [0]{.\EOS\space}%
\providecommand \EOS [0]{\spacefactor3000\relax}%
\providecommand \BibitemShut  [1]{\csname bibitem#1\endcsname}%
\let\auto@bib@innerbib\@empty
%</preamble>
\bibitem [{\citenamefont {Billy}\ \emph {et~al.}(2008)\citenamefont {Billy},
  \citenamefont {Josse}, \citenamefont {Zuo}, \citenamefont {Bernard},
  \citenamefont {Hambrecht}, \citenamefont {Lugan}, \citenamefont {Clement},
  \citenamefont {Sanchez-Palencia}, \citenamefont {Bouyer},\ and\ \citenamefont
  {Aspect}}]{Billy:2008}%
  \BibitemOpen
  \bibfield  {author} {\bibinfo {author} {\bibfnamefont {J.}~\bibnamefont
  {Billy}}, \bibinfo {author} {\bibfnamefont {V.}~\bibnamefont {Josse}},
  \bibinfo {author} {\bibfnamefont {Z.}~\bibnamefont {Zuo}}, \bibinfo {author}
  {\bibfnamefont {A.}~\bibnamefont {Bernard}}, \bibinfo {author} {\bibfnamefont
  {B.}~\bibnamefont {Hambrecht}}, \bibinfo {author} {\bibfnamefont
  {P.}~\bibnamefont {Lugan}}, \bibinfo {author} {\bibfnamefont
  {D.}~\bibnamefont {Clement}}, \bibinfo {author} {\bibfnamefont
  {L.}~\bibnamefont {Sanchez-Palencia}}, \bibinfo {author} {\bibfnamefont
  {P.}~\bibnamefont {Bouyer}}, \ and\ \bibinfo {author} {\bibfnamefont
  {A.}~\bibnamefont {Aspect}},\ }\href@noop {} {\bibfield  {journal} {\bibinfo
  {journal} {Nature}\ }\textbf {\bibinfo {volume} {453}},\ \bibinfo {pages}
  {891} (\bibinfo {year} {2008})}\BibitemShut {NoStop}%
\bibitem [{\citenamefont {Roati}\ \emph {et~al.}(2008)\citenamefont {Roati},
  \citenamefont {D'Errico}, \citenamefont {Fallani}, \citenamefont {Fattori},
  \citenamefont {Fort}, \citenamefont {Zaccanti}, \citenamefont {Modugno},
  \citenamefont {Modugno},\ and\ \citenamefont {Inguscio}}]{Roati:2008}%
  \BibitemOpen
  \bibfield  {author} {\bibinfo {author} {\bibfnamefont {G.}~\bibnamefont
  {Roati}}, \bibinfo {author} {\bibfnamefont {C.}~\bibnamefont {D'Errico}},
  \bibinfo {author} {\bibfnamefont {L.}~\bibnamefont {Fallani}}, \bibinfo
  {author} {\bibfnamefont {M.}~\bibnamefont {Fattori}}, \bibinfo {author}
  {\bibfnamefont {C.}~\bibnamefont {Fort}}, \bibinfo {author} {\bibfnamefont
  {M.}~\bibnamefont {Zaccanti}}, \bibinfo {author} {\bibfnamefont
  {G.}~\bibnamefont {Modugno}}, \bibinfo {author} {\bibfnamefont
  {M.}~\bibnamefont {Modugno}}, \ and\ \bibinfo {author} {\bibfnamefont
  {M.}~\bibnamefont {Inguscio}},\ }\href@noop {} {\bibfield  {journal}
  {\bibinfo  {journal} {Nature}\ }\textbf {\bibinfo {volume} {453}},\ \bibinfo
  {pages} {895} (\bibinfo {year} {2008})}\BibitemShut {NoStop}%
\bibitem [{\citenamefont {Levy}\ \emph {et~al.}(2007)\citenamefont {Levy},
  \citenamefont {Lahoud}, \citenamefont {Shomroni},\ and\ \citenamefont
  {Steinhauer}}]{Levy2007}%
  \BibitemOpen
  \bibfield  {author} {\bibinfo {author} {\bibfnamefont {S.}~\bibnamefont
  {Levy}}, \bibinfo {author} {\bibfnamefont {E.}~\bibnamefont {Lahoud}},
  \bibinfo {author} {\bibfnamefont {I.}~\bibnamefont {Shomroni}}, \ and\
  \bibinfo {author} {\bibfnamefont {J.}~\bibnamefont {Steinhauer}},\
  }\href@noop {} {\bibfield  {journal} {\bibinfo  {journal} {Nature}\ }\textbf
  {\bibinfo {volume} {449}},\ \bibinfo {pages} {579} (\bibinfo {year}
  {2007})}\BibitemShut {NoStop}%
\bibitem [{\citenamefont {Palzer}\ \emph {et~al.}(2009)\citenamefont {Palzer},
  \citenamefont {Zipkes}, \citenamefont {Sias},\ and\ \citenamefont
  {K{\"o}hl}}]{Kohl2009}%
  \BibitemOpen
  \bibfield  {author} {\bibinfo {author} {\bibfnamefont {S.}~\bibnamefont
  {Palzer}}, \bibinfo {author} {\bibfnamefont {C.}~\bibnamefont {Zipkes}},
  \bibinfo {author} {\bibfnamefont {C.}~\bibnamefont {Sias}}, \ and\ \bibinfo
  {author} {\bibfnamefont {M.}~\bibnamefont {K{\"o}hl}},\ }\href@noop {}
  {\bibfield  {journal} {\bibinfo  {journal} {Phys. Rev. Lett.}\ }\textbf
  {\bibinfo {volume} {103}},\ \bibinfo {pages} {150601} (\bibinfo {year}
  {2009})}\BibitemShut {NoStop}%
\bibitem [{\citenamefont {Catani}\ \emph {et~al.}(2012)\citenamefont {Catani},
  \citenamefont {Lamporesi}, \citenamefont {Naik}, \citenamefont {Gring},
  \citenamefont {Inguscio}, \citenamefont {Minardi}, \citenamefont {Kantian},\
  and\ \citenamefont {Giamarchi}}]{Minardi:12}%
  \BibitemOpen
  \bibfield  {author} {\bibinfo {author} {\bibfnamefont {J.}~\bibnamefont
  {Catani}}, \bibinfo {author} {\bibfnamefont {G.}~\bibnamefont {Lamporesi}},
  \bibinfo {author} {\bibfnamefont {D.}~\bibnamefont {Naik}}, \bibinfo {author}
  {\bibfnamefont {M.}~\bibnamefont {Gring}}, \bibinfo {author} {\bibfnamefont
  {M.}~\bibnamefont {Inguscio}}, \bibinfo {author} {\bibfnamefont
  {F.}~\bibnamefont {Minardi}}, \bibinfo {author} {\bibfnamefont
  {A.}~\bibnamefont {Kantian}}, \ and\ \bibinfo {author} {\bibfnamefont
  {T.}~\bibnamefont {Giamarchi}},\ }\href {\doibase 10.1103/PhysRevA.85.023623}
  {\bibfield  {journal} {\bibinfo  {journal} {Phys. Rev. A}\ }\textbf {\bibinfo
  {volume} {85}},\ \bibinfo {pages} {023623} (\bibinfo {year}
  {2012})}\BibitemShut {NoStop}%
\bibitem [{\citenamefont {Brantut}\ \emph {et~al.}(2012)\citenamefont
  {Brantut}, \citenamefont {Meineke}, \citenamefont {Stadler}, \citenamefont
  {Krinner},\ and\ \citenamefont {Esslinger}}]{Esslinger2}%
  \BibitemOpen
  \bibfield  {author} {\bibinfo {author} {\bibfnamefont {J.-P.}\ \bibnamefont
  {Brantut}}, \bibinfo {author} {\bibfnamefont {J.}~\bibnamefont {Meineke}},
  \bibinfo {author} {\bibfnamefont {D.}~\bibnamefont {Stadler}}, \bibinfo
  {author} {\bibfnamefont {S.}~\bibnamefont {Krinner}}, \ and\ \bibinfo
  {author} {\bibfnamefont {T.}~\bibnamefont {Esslinger}},\ }\href@noop {}
  {\bibfield  {journal} {\bibinfo  {journal} {Science}\ }\textbf {\bibinfo
  {volume} {337}},\ \bibinfo {pages} {1069} (\bibinfo {year}
  {2012})}\BibitemShut {NoStop}%
\bibitem [{\citenamefont {Stadler}\ \emph {et~al.}(2012)\citenamefont
  {Stadler}, \citenamefont {Krinner}, \citenamefont {Meineke}, \citenamefont
  {Brantut},\ and\ \citenamefont {Esslinger}}]{Esslinger1}%
  \BibitemOpen
  \bibfield  {author} {\bibinfo {author} {\bibfnamefont {D.}~\bibnamefont
  {Stadler}}, \bibinfo {author} {\bibfnamefont {S.}~\bibnamefont {Krinner}},
  \bibinfo {author} {\bibfnamefont {J.}~\bibnamefont {Meineke}}, \bibinfo
  {author} {\bibfnamefont {J.-P.}\ \bibnamefont {Brantut}}, \ and\ \bibinfo
  {author} {\bibfnamefont {T.}~\bibnamefont {Esslinger}},\ }\href@noop {}
  {\bibfield  {journal} {\bibinfo  {journal} {Nature}\ }\textbf {\bibinfo
  {volume} {491}},\ \bibinfo {pages} {736} (\bibinfo {year}
  {2012})}\BibitemShut {NoStop}%
\bibitem [{\citenamefont {Ramanathan}\ \emph {et~al.}(2011)\citenamefont
  {Ramanathan}, \citenamefont {Wright}, \citenamefont {Muniz}, \citenamefont
  {Zelan}, \citenamefont {Hill}, \citenamefont {Lobb}, \citenamefont
  {Helmerson}, \citenamefont {Phillips},\ and\ \citenamefont
  {Campbell}}]{TunWeakLink}%
  \BibitemOpen
  \bibfield  {author} {\bibinfo {author} {\bibfnamefont {A.}~\bibnamefont
  {Ramanathan}}, \bibinfo {author} {\bibfnamefont {K.~C.}\ \bibnamefont
  {Wright}}, \bibinfo {author} {\bibfnamefont {S.~R.}\ \bibnamefont {Muniz}},
  \bibinfo {author} {\bibfnamefont {M.}~\bibnamefont {Zelan}}, \bibinfo
  {author} {\bibfnamefont {W.~T.}\ \bibnamefont {Hill}}, \bibinfo {author}
  {\bibfnamefont {C.~J.}\ \bibnamefont {Lobb}}, \bibinfo {author}
  {\bibfnamefont {K.}~\bibnamefont {Helmerson}}, \bibinfo {author}
  {\bibfnamefont {W.~D.}\ \bibnamefont {Phillips}}, \ and\ \bibinfo {author}
  {\bibfnamefont {G.~K.}\ \bibnamefont {Campbell}},\ }\href {\doibase
  10.1103/PhysRevLett.106.130401} {\bibfield  {journal} {\bibinfo  {journal}
  {Phys. Rev. Lett.}\ }\textbf {\bibinfo {volume} {106}},\ \bibinfo {pages}
  {130401} (\bibinfo {year} {2011})}\BibitemShut {NoStop}%
\bibitem [{\citenamefont {Tanzi}\ \emph {et~al.}(2013)\citenamefont {Tanzi},
  \citenamefont {Lucioni}, \citenamefont {Chaudhuri}, \citenamefont {Gori},
  \citenamefont {Kumar}, \citenamefont {D'Errico}, \citenamefont {Inguscio},\
  and\ \citenamefont {Modugno}}]{Tanzi:2013}%
  \BibitemOpen
  \bibfield  {author} {\bibinfo {author} {\bibfnamefont {L.}~\bibnamefont
  {Tanzi}}, \bibinfo {author} {\bibfnamefont {E.}~\bibnamefont {Lucioni}},
  \bibinfo {author} {\bibfnamefont {S.}~\bibnamefont {Chaudhuri}}, \bibinfo
  {author} {\bibfnamefont {L.}~\bibnamefont {Gori}}, \bibinfo {author}
  {\bibfnamefont {A.}~\bibnamefont {Kumar}}, \bibinfo {author} {\bibfnamefont
  {C.}~\bibnamefont {D'Errico}}, \bibinfo {author} {\bibfnamefont
  {M.}~\bibnamefont {Inguscio}}, \ and\ \bibinfo {author} {\bibfnamefont
  {G.}~\bibnamefont {Modugno}},\ }\href@noop {} {\bibfield  {journal} {\bibinfo
   {journal} {Phys. Rev. Lett.}\ }\textbf {\bibinfo {volume} {111}},\ \bibinfo
  {pages} {115301} (\bibinfo {year} {2013})}\BibitemShut {NoStop}%
\bibitem [{\citenamefont {Cazalilla}\ \emph {et~al.}(2011)\citenamefont
  {Cazalilla}, \citenamefont {Citro}, \citenamefont {Giamarchi}, \citenamefont
  {Orignac},\ and\ \citenamefont {Rigol}}]{CazalillaRMP:11}%
  \BibitemOpen
  \bibfield  {author} {\bibinfo {author} {\bibfnamefont {M.~A.}\ \bibnamefont
  {Cazalilla}}, \bibinfo {author} {\bibfnamefont {R.}~\bibnamefont {Citro}},
  \bibinfo {author} {\bibfnamefont {T.}~\bibnamefont {Giamarchi}}, \bibinfo
  {author} {\bibfnamefont {E.}~\bibnamefont {Orignac}}, \ and\ \bibinfo
  {author} {\bibfnamefont {M.}~\bibnamefont {Rigol}},\ }\href {\doibase
  10.1103/RevModPhys.83.1405} {\bibfield  {journal} {\bibinfo  {journal} {Rev.
  Mod. Phys.}\ }\textbf {\bibinfo {volume} {83}},\ \bibinfo {pages} {1405}
  (\bibinfo {year} {2011})}\BibitemShut {NoStop}%
\bibitem [{\citenamefont {Imambekov}\ \emph {et~al.}(2012)\citenamefont
  {Imambekov}, \citenamefont {Schmidt},\ and\ \citenamefont
  {Glazman}}]{ImSchGl}%
  \BibitemOpen
  \bibfield  {author} {\bibinfo {author} {\bibfnamefont {A.}~\bibnamefont
  {Imambekov}}, \bibinfo {author} {\bibfnamefont {T.~L.}\ \bibnamefont
  {Schmidt}}, \ and\ \bibinfo {author} {\bibfnamefont {L.~I.}\ \bibnamefont
  {Glazman}},\ }\href {\doibase 10.1103/RevModPhys.84.1253} {\bibfield
  {journal} {\bibinfo  {journal} {Rev. Mod. Phys.}\ }\textbf {\bibinfo {volume}
  {84}},\ \bibinfo {pages} {1253} (\bibinfo {year} {2012})}\BibitemShut
  {NoStop}%
\bibitem [{\citenamefont {Kane}\ and\ \citenamefont
  {Fisher}(1992{\natexlab{a}})}]{KaneFis:92a}%
  \BibitemOpen
  \bibfield  {author} {\bibinfo {author} {\bibfnamefont {C.~L.}\ \bibnamefont
  {Kane}}\ and\ \bibinfo {author} {\bibfnamefont {M.~P.~A.}\ \bibnamefont
  {Fisher}},\ }\href@noop {} {\bibfield  {journal} {\bibinfo  {journal} {Phys.
  Rev. Lett.}\ }\textbf {\bibinfo {volume} {68}},\ \bibinfo {pages} {1220}
  (\bibinfo {year} {1992}{\natexlab{a}})}\BibitemShut {NoStop}%
\bibitem [{\citenamefont {Kane}\ and\ \citenamefont
  {Fisher}(1992{\natexlab{b}})}]{Kane-Fisher}%
  \BibitemOpen
  \bibfield  {author} {\bibinfo {author} {\bibfnamefont {C.~L.}\ \bibnamefont
  {Kane}}\ and\ \bibinfo {author} {\bibfnamefont {M.~P.~A.}\ \bibnamefont
  {Fisher}},\ }\href {\doibase 10.1103/PhysRevB.46.15233} {\bibfield  {journal}
  {\bibinfo  {journal} {Phys. Rev. B}\ }\textbf {\bibinfo {volume} {46}},\
  \bibinfo {pages} {15233} (\bibinfo {year} {1992}{\natexlab{b}})}\BibitemShut
  {NoStop}%
\bibitem [{\citenamefont {Matveev}\ \emph {et~al.}(1993)\citenamefont
  {Matveev}, \citenamefont {Yue},\ and\ \citenamefont
  {Glazman}}]{MatYueGlaz:93}%
  \BibitemOpen
  \bibfield  {author} {\bibinfo {author} {\bibfnamefont {K.~A.}\ \bibnamefont
  {Matveev}}, \bibinfo {author} {\bibfnamefont {D.}~\bibnamefont {Yue}}, \ and\
  \bibinfo {author} {\bibfnamefont {L.~I.}\ \bibnamefont {Glazman}},\
  }\href@noop {} {\bibfield  {journal} {\bibinfo  {journal} {Phys. Rev. Lett.}\
  }\textbf {\bibinfo {volume} {71}},\ \bibinfo {pages} {3351} (\bibinfo {year}
  {1993})}\BibitemShut {NoStop}%
\bibitem [{\citenamefont {Furusaki}\ and\ \citenamefont
  {Nagaosa}(1993)}]{FurusakiNagaosa:93b}%
  \BibitemOpen
  \bibfield  {author} {\bibinfo {author} {\bibfnamefont {A.}~\bibnamefont
  {Furusaki}}\ and\ \bibinfo {author} {\bibfnamefont {N.}~\bibnamefont
  {Nagaosa}},\ }\href {\doibase 10.1103/PhysRevB.47.4631} {\bibfield  {journal}
  {\bibinfo  {journal} {Phys. Rev. B}\ }\textbf {\bibinfo {volume} {47}},\
  \bibinfo {pages} {4631} (\bibinfo {year} {1993})}\BibitemShut {NoStop}%
\bibitem [{\citenamefont {Fabrizio}\ and\ \citenamefont
  {Gogolin}(1995)}]{FabrizioGogolin:95}%
  \BibitemOpen
  \bibfield  {author} {\bibinfo {author} {\bibfnamefont {M.}~\bibnamefont
  {Fabrizio}}\ and\ \bibinfo {author} {\bibfnamefont {A.~O.}\ \bibnamefont
  {Gogolin}},\ }\href {\doibase 10.1103/PhysRevB.51.17827} {\bibfield
  {journal} {\bibinfo  {journal} {Phys. Rev. B}\ }\textbf {\bibinfo {volume}
  {51}},\ \bibinfo {pages} {17827} (\bibinfo {year} {1995})}\BibitemShut
  {NoStop}%
\bibitem [{\citenamefont {Furusaki}\ and\ \citenamefont
  {Matveev}(2002)}]{FurMatv:02}%
  \BibitemOpen
  \bibfield  {author} {\bibinfo {author} {\bibfnamefont {A.}~\bibnamefont
  {Furusaki}}\ and\ \bibinfo {author} {\bibfnamefont {K.~A.}\ \bibnamefont
  {Matveev}},\ }\href@noop {} {\bibfield  {journal} {\bibinfo  {journal} {Phys.
  Rev. Lett.}\ }\textbf {\bibinfo {volume} {88}},\ \bibinfo {pages} {226404}
  (\bibinfo {year} {2002})}\BibitemShut {NoStop}%
\bibitem [{\citenamefont {Nazarov}\ and\ \citenamefont
  {Glazman}(2003)}]{NazGlaz:03}%
  \BibitemOpen
  \bibfield  {author} {\bibinfo {author} {\bibfnamefont {Y.~V.}\ \bibnamefont
  {Nazarov}}\ and\ \bibinfo {author} {\bibfnamefont {L.~I.}\ \bibnamefont
  {Glazman}},\ }\href@noop {} {\bibfield  {journal} {\bibinfo  {journal} {Phys.
  Rev. Lett.}\ }\textbf {\bibinfo {volume} {91}},\ \bibinfo {pages} {126804}
  (\bibinfo {year} {2003})}\BibitemShut {NoStop}%
\bibitem [{\citenamefont {Polyakov}\ and\ \citenamefont
  {Gornyi}(2003)}]{PolGorn:03}%
  \BibitemOpen
  \bibfield  {author} {\bibinfo {author} {\bibfnamefont {D.~G.}\ \bibnamefont
  {Polyakov}}\ and\ \bibinfo {author} {\bibfnamefont {I.~V.}\ \bibnamefont
  {Gornyi}},\ }\href@noop {} {\bibfield  {journal} {\bibinfo  {journal} {Phys.
  Rev. B}\ }\textbf {\bibinfo {volume} {68}},\ \bibinfo {pages} {035421}
  (\bibinfo {year} {2003})}\BibitemShut {NoStop}%
\bibitem [{\citenamefont {Lerner}\ \emph {et~al.}(2008)\citenamefont {Lerner},
  \citenamefont {Yudson},\ and\ \citenamefont {Yurkevich}}]{LYY:08}%
  \BibitemOpen
  \bibfield  {author} {\bibinfo {author} {\bibfnamefont {I.~V.}\ \bibnamefont
  {Lerner}}, \bibinfo {author} {\bibfnamefont {V.~I.}\ \bibnamefont {Yudson}},
  \ and\ \bibinfo {author} {\bibfnamefont {I.~V.}\ \bibnamefont {Yurkevich}},\
  }\href {\doibase 10.1103/PhysRevLett.100.256805} {\bibfield  {journal}
  {\bibinfo  {journal} {Phys. Rev. Lett.}\ }\textbf {\bibinfo {volume} {100}},\
  \bibinfo {pages} {256805} (\bibinfo {year} {2008})}\BibitemShut {NoStop}%
\bibitem [{\citenamefont {Goldstein}\ and\ \citenamefont
  {Berkovits}(2010)}]{GB:10}%
  \BibitemOpen
  \bibfield  {author} {\bibinfo {author} {\bibfnamefont {M.}~\bibnamefont
  {Goldstein}}\ and\ \bibinfo {author} {\bibfnamefont {R.}~\bibnamefont
  {Berkovits}},\ }\href {\doibase 10.1103/PhysRevLett.104.106403} {\bibfield
  {journal} {\bibinfo  {journal} {Phys. Rev. Lett.}\ }\textbf {\bibinfo
  {volume} {104}},\ \bibinfo {pages} {106403} (\bibinfo {year}
  {2010})}\BibitemShut {NoStop}%
\bibitem [{\citenamefont {Bockrath}\ \emph {et~al.}(1999)\citenamefont
  {Bockrath}, \citenamefont {Cobden}, \citenamefont {Lu}, \citenamefont
  {Rinzler}, \citenamefont {Smalley}, \citenamefont {Balents},\ and\
  \citenamefont {McEuen}}]{Bockrath:99}%
  \BibitemOpen
  \bibfield  {author} {\bibinfo {author} {\bibfnamefont {M.}~\bibnamefont
  {Bockrath}}, \bibinfo {author} {\bibfnamefont {D.~H.}\ \bibnamefont
  {Cobden}}, \bibinfo {author} {\bibfnamefont {J.}~\bibnamefont {Lu}}, \bibinfo
  {author} {\bibfnamefont {A.~G.}\ \bibnamefont {Rinzler}}, \bibinfo {author}
  {\bibfnamefont {R.~E.}\ \bibnamefont {Smalley}}, \bibinfo {author}
  {\bibfnamefont {L.}~\bibnamefont {Balents}}, \ and\ \bibinfo {author}
  {\bibfnamefont {P.~L.}\ \bibnamefont {McEuen}},\ }\href@noop {} {\bibfield
  {journal} {\bibinfo  {journal} {Nature}\ }\textbf {\bibinfo {volume} {397}},\
  \bibinfo {pages} {598} (\bibinfo {year} {1999})}\BibitemShut {NoStop}%
\bibitem [{\citenamefont {Bockrath}\ \emph {et~al.}(2001)\citenamefont
  {Bockrath}, \citenamefont {Liang}, \citenamefont {Bozovic}, \citenamefont
  {Hafner}, \citenamefont {Lieber}, \citenamefont {Tinkham},\ and\
  \citenamefont {Park}}]{Bockrath:01}%
  \BibitemOpen
  \bibfield  {author} {\bibinfo {author} {\bibfnamefont {M.}~\bibnamefont
  {Bockrath}}, \bibinfo {author} {\bibfnamefont {W.~J.}\ \bibnamefont {Liang}},
  \bibinfo {author} {\bibfnamefont {D.}~\bibnamefont {Bozovic}}, \bibinfo
  {author} {\bibfnamefont {J.~H.}\ \bibnamefont {Hafner}}, \bibinfo {author}
  {\bibfnamefont {C.~M.}\ \bibnamefont {Lieber}}, \bibinfo {author}
  {\bibfnamefont {M.}~\bibnamefont {Tinkham}}, \ and\ \bibinfo {author}
  {\bibfnamefont {H.~K.}\ \bibnamefont {Park}},\ }\href@noop {} {\bibfield
  {journal} {\bibinfo  {journal} {Science}\ }\textbf {\bibinfo {volume}
  {291}},\ \bibinfo {pages} {283} (\bibinfo {year} {2001})}\BibitemShut
  {NoStop}%
\bibitem [{\citenamefont {Yao}\ \emph {et~al.}(1999)\citenamefont {Yao},
  \citenamefont {Postma}, \citenamefont {Balents},\ and\ \citenamefont
  {Dekker}}]{Yao:99}%
  \BibitemOpen
  \bibfield  {author} {\bibinfo {author} {\bibfnamefont {Z.}~\bibnamefont
  {Yao}}, \bibinfo {author} {\bibfnamefont {H.~W.~C.}\ \bibnamefont {Postma}},
  \bibinfo {author} {\bibfnamefont {L.}~\bibnamefont {Balents}}, \ and\
  \bibinfo {author} {\bibfnamefont {C.}~\bibnamefont {Dekker}},\ }\href@noop {}
  {\bibfield  {journal} {\bibinfo  {journal} {Nature}\ }\textbf {\bibinfo
  {volume} {402}},\ \bibinfo {pages} {273} (\bibinfo {year}
  {1999})}\BibitemShut {NoStop}%
\bibitem [{\citenamefont {Auslaender}\ \emph {et~al.}(2002)\citenamefont
  {Auslaender}, \citenamefont {Yacoby}, \citenamefont {de~Picciotto},
  \citenamefont {Baldwin}, \citenamefont {Pfeiffer},\ and\ \citenamefont
  {West}}]{Auslaender:02}%
  \BibitemOpen
  \bibfield  {author} {\bibinfo {author} {\bibfnamefont {O.~M.}\ \bibnamefont
  {Auslaender}}, \bibinfo {author} {\bibfnamefont {A.}~\bibnamefont {Yacoby}},
  \bibinfo {author} {\bibfnamefont {R.}~\bibnamefont {de~Picciotto}}, \bibinfo
  {author} {\bibfnamefont {K.~W.}\ \bibnamefont {Baldwin}}, \bibinfo {author}
  {\bibfnamefont {L.~N.}\ \bibnamefont {Pfeiffer}}, \ and\ \bibinfo {author}
  {\bibfnamefont {K.~W.}\ \bibnamefont {West}},\ }\href@noop {} {\bibfield
  {journal} {\bibinfo  {journal} {Science}\ }\textbf {\bibinfo {volume}
  {295}},\ \bibinfo {pages} {825} (\bibinfo {year} {2002})}\BibitemShut
  {NoStop}%
\bibitem [{\citenamefont {Venkataraman}\ \emph {et~al.}(2006)\citenamefont
  {Venkataraman}, \citenamefont {Hong},\ and\ \citenamefont {Kim}}]{Kim:06}%
  \BibitemOpen
  \bibfield  {author} {\bibinfo {author} {\bibfnamefont {L.}~\bibnamefont
  {Venkataraman}}, \bibinfo {author} {\bibfnamefont {Y.~S.}\ \bibnamefont
  {Hong}}, \ and\ \bibinfo {author} {\bibfnamefont {P.}~\bibnamefont {Kim}},\
  }\href {\doibase 10.1103/PhysRevLett.96.076601} {\bibfield  {journal}
  {\bibinfo  {journal} {Phys. Rev. Lett.}\ }\textbf {\bibinfo {volume} {96}},\
  \bibinfo {pages} {076601} (\bibinfo {year} {2006})}\BibitemShut {NoStop}%
\bibitem [{\citenamefont {Levy}\ \emph {et~al.}(2006)\citenamefont {Levy},
  \citenamefont {Tsukernik}, \citenamefont {Karpovski}, \citenamefont
  {Palevski}, \citenamefont {Dwir}, \citenamefont {Pelucchi}, \citenamefont
  {Rudra}, \citenamefont {Kapon},\ and\ \citenamefont {Oreg}}]{Levy:06}%
  \BibitemOpen
  \bibfield  {author} {\bibinfo {author} {\bibfnamefont {E.}~\bibnamefont
  {Levy}}, \bibinfo {author} {\bibfnamefont {A.}~\bibnamefont {Tsukernik}},
  \bibinfo {author} {\bibfnamefont {M.}~\bibnamefont {Karpovski}}, \bibinfo
  {author} {\bibfnamefont {A.}~\bibnamefont {Palevski}}, \bibinfo {author}
  {\bibfnamefont {B.}~\bibnamefont {Dwir}}, \bibinfo {author} {\bibfnamefont
  {E.}~\bibnamefont {Pelucchi}}, \bibinfo {author} {\bibfnamefont
  {A.}~\bibnamefont {Rudra}}, \bibinfo {author} {\bibfnamefont
  {E.}~\bibnamefont {Kapon}}, \ and\ \bibinfo {author} {\bibfnamefont
  {Y.}~\bibnamefont {Oreg}},\ }\href {\doibase 10.1103/PhysRevLett.97.196802}
  {\bibfield  {journal} {\bibinfo  {journal} {Phys. Rev. Lett.}\ }\textbf
  {\bibinfo {volume} {97}},\ \bibinfo {pages} {196802} (\bibinfo {year}
  {2006})}\BibitemShut {NoStop}%
\bibitem [{\citenamefont {Levy}\ \emph {et~al.}(2012)\citenamefont {Levy},
  \citenamefont {Sternfeld}, \citenamefont {Eshkol}, \citenamefont {Karpovski},
  \citenamefont {Dwir}, \citenamefont {Rudra}, \citenamefont {Kapon},
  \citenamefont {Oreg},\ and\ \citenamefont {Palevski}}]{Levy:2012}%
  \BibitemOpen
  \bibfield  {author} {\bibinfo {author} {\bibfnamefont {E.}~\bibnamefont
  {Levy}}, \bibinfo {author} {\bibfnamefont {I.}~\bibnamefont {Sternfeld}},
  \bibinfo {author} {\bibfnamefont {M.}~\bibnamefont {Eshkol}}, \bibinfo
  {author} {\bibfnamefont {M.}~\bibnamefont {Karpovski}}, \bibinfo {author}
  {\bibfnamefont {B.}~\bibnamefont {Dwir}}, \bibinfo {author} {\bibfnamefont
  {A.}~\bibnamefont {Rudra}}, \bibinfo {author} {\bibfnamefont
  {E.}~\bibnamefont {Kapon}}, \bibinfo {author} {\bibfnamefont
  {Y.}~\bibnamefont {Oreg}}, \ and\ \bibinfo {author} {\bibfnamefont
  {A.}~\bibnamefont {Palevski}},\ }\href {\doibase 10.1103/PhysRevB.85.045315}
  {\bibfield  {journal} {\bibinfo  {journal} {Phys. Rev. B}\ }\textbf {\bibinfo
  {volume} {85}},\ \bibinfo {pages} {045315} (\bibinfo {year}
  {2012})}\BibitemShut {NoStop}%
\bibitem [{\citenamefont {Paul}\ \emph {et~al.}(2007)\citenamefont {Paul},
  \citenamefont {Hartung}, \citenamefont {Richter},\ and\ \citenamefont
  {Schlagheck}}]{Paul2007}%
  \BibitemOpen
  \bibfield  {author} {\bibinfo {author} {\bibfnamefont {T.}~\bibnamefont
  {Paul}}, \bibinfo {author} {\bibfnamefont {M.}~\bibnamefont {Hartung}},
  \bibinfo {author} {\bibfnamefont {K.}~\bibnamefont {Richter}}, \ and\
  \bibinfo {author} {\bibfnamefont {P.}~\bibnamefont {Schlagheck}},\ }\href
  {\doibase 10.1103/PhysRevA.76.063605} {\bibfield  {journal} {\bibinfo
  {journal} {Phys. Rev. A}\ }\textbf {\bibinfo {volume} {76}},\ \bibinfo
  {pages} {063605} (\bibinfo {year} {2007})}\BibitemShut {NoStop}%
\bibitem [{\citenamefont {Gutman}\ \emph {et~al.}(2012)\citenamefont {Gutman},
  \citenamefont {Gefen},\ and\ \citenamefont
  {Mirlin}}]{Gutman-Gefen-Mirlin2012}%
  \BibitemOpen
  \bibfield  {author} {\bibinfo {author} {\bibfnamefont {D.~B.}\ \bibnamefont
  {Gutman}}, \bibinfo {author} {\bibfnamefont {Y.}~\bibnamefont {Gefen}}, \
  and\ \bibinfo {author} {\bibfnamefont {A.~D.}\ \bibnamefont {Mirlin}},\
  }\href {\doibase 10.1103/PhysRevB.85.125102} {\bibfield  {journal} {\bibinfo
  {journal} {Phys. Rev. B}\ }\textbf {\bibinfo {volume} {85}},\ \bibinfo
  {pages} {125102} (\bibinfo {year} {2012})}\BibitemShut {NoStop}%
\bibitem [{\citenamefont {Kristinsd\'ottir}\ \emph {et~al.}(2013)\citenamefont
  {Kristinsd\'ottir}, \citenamefont {Karlstr\"om}, \citenamefont {Bjerlin},
  \citenamefont {Cremon}, \citenamefont {Schlagheck}, \citenamefont {Wacker},\
  and\ \citenamefont {Reimann}}]{Reimann2013}%
  \BibitemOpen
  \bibfield  {author} {\bibinfo {author} {\bibfnamefont {L.~H.}\ \bibnamefont
  {Kristinsd\'ottir}}, \bibinfo {author} {\bibfnamefont {O.}~\bibnamefont
  {Karlstr\"om}}, \bibinfo {author} {\bibfnamefont {J.}~\bibnamefont
  {Bjerlin}}, \bibinfo {author} {\bibfnamefont {J.~C.}\ \bibnamefont {Cremon}},
  \bibinfo {author} {\bibfnamefont {P.}~\bibnamefont {Schlagheck}}, \bibinfo
  {author} {\bibfnamefont {A.}~\bibnamefont {Wacker}}, \ and\ \bibinfo {author}
  {\bibfnamefont {S.~M.}\ \bibnamefont {Reimann}},\ }\href {\doibase
  10.1103/PhysRevLett.110.085303} {\bibfield  {journal} {\bibinfo  {journal}
  {Phys. Rev. Lett.}\ }\textbf {\bibinfo {volume} {110}},\ \bibinfo {pages}
  {085303} (\bibinfo {year} {2013})}\BibitemShut {NoStop}%
\bibitem [{\citenamefont {Fazio}\ \emph {et~al.}(1995)\citenamefont {Fazio},
  \citenamefont {Hekking},\ and\ \citenamefont {Odintsov}}]{Fazio1}%
  \BibitemOpen
  \bibfield  {author} {\bibinfo {author} {\bibfnamefont {R.}~\bibnamefont
  {Fazio}}, \bibinfo {author} {\bibfnamefont {F.~W.~J.}\ \bibnamefont
  {Hekking}}, \ and\ \bibinfo {author} {\bibfnamefont {A.~A.}\ \bibnamefont
  {Odintsov}},\ }\href {\doibase 10.1103/PhysRevLett.74.1843} {\bibfield
  {journal} {\bibinfo  {journal} {Phys. Rev. Lett.}\ }\textbf {\bibinfo
  {volume} {74}},\ \bibinfo {pages} {1843} (\bibinfo {year}
  {1995})}\BibitemShut {NoStop}%
\bibitem [{\citenamefont {Fazio}\ \emph {et~al.}(1996)\citenamefont {Fazio},
  \citenamefont {Hekking},\ and\ \citenamefont {Odintsov}}]{Fazio}%
  \BibitemOpen
  \bibfield  {author} {\bibinfo {author} {\bibfnamefont {R.}~\bibnamefont
  {Fazio}}, \bibinfo {author} {\bibfnamefont {F.~W.~J.}\ \bibnamefont
  {Hekking}}, \ and\ \bibinfo {author} {\bibfnamefont {A.~A.}\ \bibnamefont
  {Odintsov}},\ }\href {\doibase 10.1103/PhysRevB.53.6653} {\bibfield
  {journal} {\bibinfo  {journal} {Phys. Rev. B}\ }\textbf {\bibinfo {volume}
  {53}},\ \bibinfo {pages} {6653} (\bibinfo {year} {1996})}\BibitemShut
  {NoStop}%
\bibitem [{\citenamefont {Maslov}\ \emph {et~al.}(1996)\citenamefont {Maslov},
  \citenamefont {Stone}, \citenamefont {Goldbart},\ and\ \citenamefont
  {Loss}}]{MasStoneGoldbartLoss}%
  \BibitemOpen
  \bibfield  {author} {\bibinfo {author} {\bibfnamefont {D.~L.}\ \bibnamefont
  {Maslov}}, \bibinfo {author} {\bibfnamefont {M.}~\bibnamefont {Stone}},
  \bibinfo {author} {\bibfnamefont {P.~M.}\ \bibnamefont {Goldbart}}, \ and\
  \bibinfo {author} {\bibfnamefont {D.}~\bibnamefont {Loss}},\ }\href {\doibase
  10.1103/PhysRevB.53.1548} {\bibfield  {journal} {\bibinfo  {journal} {Phys.
  Rev. B}\ }\textbf {\bibinfo {volume} {53}},\ \bibinfo {pages} {1548}
  (\bibinfo {year} {1996})}\BibitemShut {NoStop}%
\bibitem [{\citenamefont {Affleck}\ \emph {et~al.}(2000)\citenamefont
  {Affleck}, \citenamefont {Caux},\ and\ \citenamefont
  {Zagoskin}}]{AffleckCauxZ}%
  \BibitemOpen
  \bibfield  {author} {\bibinfo {author} {\bibfnamefont {I.}~\bibnamefont
  {Affleck}}, \bibinfo {author} {\bibfnamefont {J.-S.}\ \bibnamefont {Caux}}, \
  and\ \bibinfo {author} {\bibfnamefont {A.~M.}\ \bibnamefont {Zagoskin}},\
  }\href {\doibase 10.1103/PhysRevB.62.1433} {\bibfield  {journal} {\bibinfo
  {journal} {Phys. Rev. B}\ }\textbf {\bibinfo {volume} {62}},\ \bibinfo
  {pages} {1433} (\bibinfo {year} {2000})}\BibitemShut {NoStop}%
\bibitem [{\citenamefont {Caux}\ \emph {et~al.}(2002)\citenamefont {Caux},
  \citenamefont {Saleur},\ and\ \citenamefont {Siano}}]{Caux1}%
  \BibitemOpen
  \bibfield  {author} {\bibinfo {author} {\bibfnamefont {J.-S.}\ \bibnamefont
  {Caux}}, \bibinfo {author} {\bibfnamefont {H.}~\bibnamefont {Saleur}}, \ and\
  \bibinfo {author} {\bibfnamefont {F.}~\bibnamefont {Siano}},\ }\href
  {\doibase 10.1103/PhysRevLett.88.106402} {\bibfield  {journal} {\bibinfo
  {journal} {Phys. Rev. Lett.}\ }\textbf {\bibinfo {volume} {88}},\ \bibinfo
  {pages} {106402} (\bibinfo {year} {2002})}\BibitemShut {NoStop}%
\bibitem [{\citenamefont {Caux}\ \emph {et~al.}(2003)\citenamefont {Caux},
  \citenamefont {Saleur},\ and\ \citenamefont {Siano}}]{Caux2}%
  \BibitemOpen
  \bibfield  {author} {\bibinfo {author} {\bibfnamefont {J.-S.}\ \bibnamefont
  {Caux}}, \bibinfo {author} {\bibfnamefont {H.}~\bibnamefont {Saleur}}, \ and\
  \bibinfo {author} {\bibfnamefont {F.}~\bibnamefont {Siano}},\ }\href
  {\doibase 10.1016/j.nuclphysb.2003.08.039} {\bibfield  {journal} {\bibinfo
  {journal} {Nucl. Phys. B}\ }\textbf {\bibinfo {volume} {672}},\ \bibinfo
  {pages} {411} (\bibinfo {year} {2003})}\BibitemShut {NoStop}%
\bibitem [{\citenamefont {Folman}\ \emph {et~al.}(2000)\citenamefont {Folman},
  \citenamefont {Kr\"uger}, \citenamefont {Cassettari}, \citenamefont {Hessmo},
  \citenamefont {Maier},\ and\ \citenamefont
  {Schmiedmayer}}]{Folman-Kruger-et-al}%
  \BibitemOpen
  \bibfield  {author} {\bibinfo {author} {\bibfnamefont {R.}~\bibnamefont
  {Folman}}, \bibinfo {author} {\bibfnamefont {P.}~\bibnamefont {Kr\"uger}},
  \bibinfo {author} {\bibfnamefont {D.}~\bibnamefont {Cassettari}}, \bibinfo
  {author} {\bibfnamefont {B.}~\bibnamefont {Hessmo}}, \bibinfo {author}
  {\bibfnamefont {T.}~\bibnamefont {Maier}}, \ and\ \bibinfo {author}
  {\bibfnamefont {J.}~\bibnamefont {Schmiedmayer}},\ }\href {\doibase
  10.1103/PhysRevLett.84.4749} {\bibfield  {journal} {\bibinfo  {journal}
  {Phys. Rev. Lett.}\ }\textbf {\bibinfo {volume} {84}},\ \bibinfo {pages}
  {4749} (\bibinfo {year} {2000})}\BibitemShut {NoStop}%
\bibitem [{\citenamefont {Folman}\ \emph {et~al.}(2002)\citenamefont {Folman},
  \citenamefont {Kr\"uger}, \citenamefont {Schmiedmayer}, \citenamefont
  {Denschlag},\ and\ \citenamefont {Henkel}}]{FKS02}%
  \BibitemOpen
  \bibfield  {author} {\bibinfo {author} {\bibfnamefont {R.}~\bibnamefont
  {Folman}}, \bibinfo {author} {\bibfnamefont {P.}~\bibnamefont {Kr\"uger}},
  \bibinfo {author} {\bibfnamefont {J.}~\bibnamefont {Schmiedmayer}}, \bibinfo
  {author} {\bibfnamefont {J.}~\bibnamefont {Denschlag}}, \ and\ \bibinfo
  {author} {\bibfnamefont {C.}~\bibnamefont {Henkel}},\ }\href {\doibase
  http://dx.doi.org/10.1016/S1049-250X(02)80011-8} {\ \bibinfo {series}
  {Advances In Atomic, Molecular, and Optical Physics},\ \textbf {\bibinfo
  {volume} {48}},\ \bibinfo {pages} {263356} (\bibinfo {year}
  {2002})}\BibitemShut {NoStop}%
\bibitem [{SGL()}]{SGLK:1}%
  \BibitemOpen
  \href@noop {} {}\bibinfo {note} {{S}ee \emph{Supplemental Online Materials}
  for detail.}\BibitemShut {Stop}%
\bibitem [{\citenamefont {Giamarchi}(2004)}]{Giamarchi}%
  \BibitemOpen
  \bibfield  {author} {\bibinfo {author} {\bibfnamefont {T.}~\bibnamefont
  {Giamarchi}},\ }\href@noop {} {\emph {\bibinfo {title} {Quantum Physics in
  One Dimension}}}\ (\bibinfo  {publisher} {Clarendon Press},\ \bibinfo
  {address} {London},\ \bibinfo {year} {2004})\BibitemShut {NoStop}%
\bibitem [{\citenamefont {Gogolin}\ \emph {et~al.}(2004)\citenamefont
  {Gogolin}, \citenamefont {Nersesyan},\ and\ \citenamefont
  {Tsvelik}}]{GogNersTsv}%
  \BibitemOpen
  \bibfield  {author} {\bibinfo {author} {\bibfnamefont {A.~O.}\ \bibnamefont
  {Gogolin}}, \bibinfo {author} {\bibfnamefont {A.~A.}\ \bibnamefont
  {Nersesyan}}, \ and\ \bibinfo {author} {\bibfnamefont {A.~M.}\ \bibnamefont
  {Tsvelik}},\ }\href@noop {} {\emph {\bibinfo {title} {Bosonization and
  Strongly Correlated Systems}}}\ (\bibinfo  {publisher} {Cambridge University
  Press},\ \bibinfo {address} {Cambridge},\ \bibinfo {year} {2004})\BibitemShut
  {NoStop}%
\bibitem [{Note1()}]{Note1}%
  \BibitemOpen
  \bibinfo {note} {Here and elsewhere in the Letter we use units with $\hbar
  =1$.}\BibitemShut {Stop}%
\bibitem [{Note2()}]{Note2}%
  \BibitemOpen
  \bibinfo {note} {Allowing for a difference in the tunneling energies at both
  barriers leads insignificant changes in parameters without qualitatively
  affecting our results \cite {SGLK:1}.}\BibitemShut {Stop}%
\bibitem [{\citenamefont {Tinkham}(1996)}]{Tinkham}%
  \BibitemOpen
  \bibfield  {author} {\bibinfo {author} {\bibfnamefont {M.}~\bibnamefont
  {Tinkham}},\ }\href@noop {} {\emph {\bibinfo {title} {Introduction to
  Superconductivity}}}\ (\bibinfo  {publisher} {Dover},\ \bibinfo {address}
  {New York},\ \bibinfo {year} {1996})\ \bibinfo {note} {p.226}\BibitemShut
  {NoStop}%
\bibitem [{\citenamefont {B\"uchler}\ \emph {et~al.}(2001)\citenamefont
  {B\"uchler}, \citenamefont {Geshkenbein},\ and\ \citenamefont
  {Blatter}}]{Buchler2001}%
  \BibitemOpen
  \bibfield  {author} {\bibinfo {author} {\bibfnamefont {H.~P.}\ \bibnamefont
  {B\"uchler}}, \bibinfo {author} {\bibfnamefont {V.~B.}\ \bibnamefont
  {Geshkenbein}}, \ and\ \bibinfo {author} {\bibfnamefont {G.}~\bibnamefont
  {Blatter}},\ }\href {\doibase 10.1103/PhysRevLett.87.100403} {\bibfield
  {journal} {\bibinfo  {journal} {Phys. Rev. Lett.}\ }\textbf {\bibinfo
  {volume} {87}},\ \bibinfo {pages} {100403} (\bibinfo {year}
  {2001})}\BibitemShut {NoStop}%
\bibitem [{\citenamefont {Schecter}\ \emph {et~al.}(2012)\citenamefont
  {Schecter}, \citenamefont {Kamenev}, \citenamefont {Gangardt},\ and\
  \citenamefont {Lamacraft}}]{Schecter2012}%
  \BibitemOpen
  \bibfield  {author} {\bibinfo {author} {\bibfnamefont {M.}~\bibnamefont
  {Schecter}}, \bibinfo {author} {\bibfnamefont {A.}~\bibnamefont {Kamenev}},
  \bibinfo {author} {\bibfnamefont {D.~M.}\ \bibnamefont {Gangardt}}, \ and\
  \bibinfo {author} {\bibfnamefont {A.}~\bibnamefont {Lamacraft}},\ }\href
  {\doibase 10.1103/PhysRevLett.108.207001} {\bibfield  {journal} {\bibinfo
  {journal} {Phys. Rev. Lett.}\ }\textbf {\bibinfo {volume} {108}},\ \bibinfo
  {pages} {207001} (\bibinfo {year} {2012})}\BibitemShut {NoStop}%
\end{thebibliography}

\begin{thebibliography}{3}%
\makeatletter
\providecommand \@ifxundefined [1]{%
 \@ifx{#1\undefined}
}%
\providecommand \@ifnum [1]{%
 \ifnum #1\expandafter \@firstoftwo
 \else \expandafter \@secondoftwo
 \fi
}%
\providecommand \@ifx [1]{%
 \ifx #1\expandafter \@firstoftwo
 \else \expandafter \@secondoftwo
 \fi
}%
\providecommand \natexlab [1]{#1}%
\providecommand \enquote  [1]{``#1''}%
\providecommand \bibnamefont  [1]{#1}%
\providecommand \bibfnamefont [1]{#1}%
\providecommand \citenamefont [1]{#1}%
\providecommand \href@noop [0]{\@secondoftwo}%
\providecommand \href [0]{\begingroup \@sanitize@url \@href}%
\providecommand \@href[1]{\@@startlink{#1}\@@href}%
\providecommand \@@href[1]{\endgroup#1\@@endlink}%
\providecommand \@sanitize@url [0]{\catcode `\\12\catcode `\$12\catcode
  `\&12\catcode `\#12\catcode `\^12\catcode `\_12\catcode `\%12\relax}%
\providecommand \@@startlink[1]{}%
\providecommand \@@endlink[0]{}%
\providecommand \url  [0]{\begingroup\@sanitize@url \@url }%
\providecommand \@url [1]{\endgroup\@href {#1}{\urlprefix }}%
\providecommand \urlprefix  [0]{URL }%
\providecommand \Eprint [0]{\href }%
\providecommand \doibase [0]{http://dx.doi.org/}%
\providecommand \selectlanguage [0]{\@gobble}%
\providecommand \bibinfo  [0]{\@secondoftwo}%
\providecommand \bibfield  [0]{\@secondoftwo}%
\providecommand \translation [1]{[#1]}%
\providecommand \BibitemOpen [0]{}%
\providecommand \bibitemStop [0]{}%
\providecommand \bibitemNoStop [0]{.\EOS\space}%
\providecommand \EOS [0]{\spacefactor3000\relax}%
\providecommand \BibitemShut  [1]{\csname bibitem#1\endcsname}%
\let\auto@bib@innerbib\@empty
%</preamble>
\bibitem [{\citenamefont {Smith}\ \emph {et~al.}(2011)\citenamefont {Smith},
  \citenamefont {Aigner}, \citenamefont {Hofferberth}, \citenamefont {Gring},
  \citenamefont {Andersson}, \citenamefont {Wildermuth}, \citenamefont
  {Kr\"{u}ger}, \citenamefont {Schneider}, \citenamefont {Schumm},\ and\
  \citenamefont {Schmiedmayer}}]{Smith11}%
  \BibitemOpen
  \bibfield  {author} {\bibinfo {author} {\bibfnamefont {D.~A.}\ \bibnamefont
  {Smith}}, \bibinfo {author} {\bibfnamefont {S.}~\bibnamefont {Aigner}},
  \bibinfo {author} {\bibfnamefont {S.}~\bibnamefont {Hofferberth}}, \bibinfo
  {author} {\bibfnamefont {M.}~\bibnamefont {Gring}}, \bibinfo {author}
  {\bibfnamefont {M.}~\bibnamefont {Andersson}}, \bibinfo {author}
  {\bibfnamefont {S.}~\bibnamefont {Wildermuth}}, \bibinfo {author}
  {\bibfnamefont {P.}~\bibnamefont {Kr\"{u}ger}}, \bibinfo {author}
  {\bibfnamefont {S.}~\bibnamefont {Schneider}}, \bibinfo {author}
  {\bibfnamefont {T.}~\bibnamefont {Schumm}}, \ and\ \bibinfo {author}
  {\bibfnamefont {J.}~\bibnamefont {Schmiedmayer}},\ }\href@noop {} {\bibfield
  {journal} {\bibinfo  {journal} {Opt. Express}\ }\textbf {\bibinfo {volume}
  {19}},\ \bibinfo {pages} {8471} (\bibinfo {year} {2011})}\BibitemShut
  {NoStop}%
\bibitem [{\citenamefont {Buecker}\ \emph {et~al.}(2009)\citenamefont
  {Buecker}, \citenamefont {Perrin}, \citenamefont {Manz}, \citenamefont
  {Betz}, \citenamefont {Koller}, \citenamefont {Plisson}, \citenamefont
  {Rottmann}, \citenamefont {Schumm},\ and\ \citenamefont
  {Schmiedmayer}}]{Plisson}%
  \BibitemOpen
  \bibfield  {author} {\bibinfo {author} {\bibfnamefont {R.}~\bibnamefont
  {Buecker}}, \bibinfo {author} {\bibfnamefont {A.}~\bibnamefont {Perrin}},
  \bibinfo {author} {\bibfnamefont {S.}~\bibnamefont {Manz}}, \bibinfo {author}
  {\bibfnamefont {T.}~\bibnamefont {Betz}}, \bibinfo {author} {\bibfnamefont
  {C.}~\bibnamefont {Koller}}, \bibinfo {author} {\bibfnamefont
  {T.}~\bibnamefont {Plisson}}, \bibinfo {author} {\bibfnamefont
  {J.}~\bibnamefont {Rottmann}}, \bibinfo {author} {\bibfnamefont
  {T.}~\bibnamefont {Schumm}}, \ and\ \bibinfo {author} {\bibfnamefont
  {J.}~\bibnamefont {Schmiedmayer}},\ }\href@noop {} {\bibfield  {journal}
  {\bibinfo  {journal} {New J. Phys.}\ }\textbf {\bibinfo {volume} {11}},\
  \bibinfo {pages} {103039} (\bibinfo {year} {2009})}\BibitemShut {NoStop}%
\bibitem [{\citenamefont {Schumm}\ \emph {et~al.}(2005)\citenamefont {Schumm},
  \citenamefont {Hofferberth}, \citenamefont {Andersson}, \citenamefont
  {Wildermuth}, \citenamefont {Groth}, \citenamefont {Bar-Joseph},
  \citenamefont {Schmiedmayer},\ and\ \citenamefont {Kr\"{u}ger}}]{Schumm2005}%
  \BibitemOpen
  \bibfield  {author} {\bibinfo {author} {\bibfnamefont {T.}~\bibnamefont
  {Schumm}}, \bibinfo {author} {\bibfnamefont {S.}~\bibnamefont {Hofferberth}},
  \bibinfo {author} {\bibfnamefont {L.~M.}\ \bibnamefont {Andersson}}, \bibinfo
  {author} {\bibfnamefont {S.}~\bibnamefont {Wildermuth}}, \bibinfo {author}
  {\bibfnamefont {S.}~\bibnamefont {Groth}}, \bibinfo {author} {\bibfnamefont
  {I.}~\bibnamefont {Bar-Joseph}}, \bibinfo {author} {\bibfnamefont
  {J.}~\bibnamefont {Schmiedmayer}}, \ and\ \bibinfo {author} {\bibfnamefont
  {P.}~\bibnamefont {Kr\"{u}ger}},\ }\href {\doibase 10.1038/nphys125}
  {\bibfield  {journal} {\bibinfo  {journal} {Nat. Phys.}\ }\textbf {\bibinfo
  {volume} {1}},\ \bibinfo {pages} {57–62} (\bibinfo {year}
  {2005})}\BibitemShut {NoStop}%
\end{thebibliography}
\end{document}